\documentclass[final,5p,times,twocolumn,authoryear]{elsarticle}

\usepackage{amssymb}
\usepackage{amsmath}
\usepackage{bm}
\usepackage{subfigure}

\usepackage[intoc]{nomencl}
\usepackage{ifthen}
\makenomenclature
\newcommand{\nomunit}[1]{\hfill\mbox{[#1]}}

\renewcommand{\nomgroup}[1]{%
  \ifthenelse{\equal{#1}{A}}{\item[\bfseries Latin symbols]}{%
  \ifthenelse{\equal{#1}{G}}{\item[\bfseries Greek symbols]}{%
  \ifthenelse{\equal{#1}{S}}{\item[\bfseries Subscripts]}{}}}%
}

\usepackage{booktabs}
\usepackage[hidelinks]{hyperref}
\usepackage{siunitx}
\begin{document}

\begin{frontmatter}

\title{Influence of fluidizing medium on hydrodynamics and particle mixing in a binary fluidized bed: a CFD--DEM study}

\author[aff1]{Ravinder Nath}

\author[aff1]{Gaurav Bhutani\corref{cor1}}
\ead{gaurav@iitmandi.ac.in}

\cortext[cor1]{Corresponding author}

\affiliation[aff1]{
	organization={School of Mechanical and Materials Engineering, Indian Institute of Technology Mandi},
	city={Mandi},
	postcode={175075},
	state={Himachal Pradesh},
	country={India}
}

\begin{abstract}
Three-dimensional CFD--DEM simulations were used to investigate the
influence of fluidizing medium on the hydrodynamics and mixing of a binary fluidized bed containing equal-density 3 and 4~mm particles. Water and air were compared using identical geometry, particle properties, and initial conditions over matched normalized superficial velocities of $0.5$--$3.0$ times the minimum fluidization velocity. The numerical framework was validated against published liquid--solid bed-expansion data and benchmarked against an established gas--solid CFD--DEM case. At matched normalized superficial velocities, water fluidization produced a comparatively dense and homogeneous bed with stable bed height and pressure drop and weak velocity fluctuations. Air fluidization resulted in greater expansion, persistent oscillations, void- and bubble-like structures, stronger circulation, and greater spatial heterogeneity. Mixing, quantified using the Lacey mixing index, increased with superficial velocity in both media but developed more rapidly and reached higher levels in air. Particle trajectories and mean-square displacement further showed greater particle mobility and transport in air. The results demonstrate that normalization by the minimum fluidization velocity does not collapse the hydrodynamic or mixing behavior across fluidizing media because fluid properties strongly influence the underlying particle-transport mechanisms.
\end{abstract}


\end{frontmatter}

\section{Introduction}
\label{sec1}

Fluidized beds are widely used in industries such as mineral processing, food processing, pharmaceuticals, and petrochemical processing. Their widespread use stems from several distinctive advantages, including the absence of moving mechanical parts, minimal shear zones, efficient fluid--solid contact, and enhanced heat and mass transfer rates. Compared with conventional mechanically agitated systems, these features translate to simpler operation and reduced maintenance costs. As a result, fluidized beds are widely used for large-scale physical and chemical processing. However, their design, operation, and scale-up demand a thorough understanding of particle dynamics and multiphase hydrodynamics.

In practical systems, fluidized beds often involve particles that differ in size and shape \cite{zhou2022cfd, zhang2021numerical}, and are fluidized using different media such as air \cite{bai2023gas} or water \cite{xie2025cfd}. 
Particle size distribution strongly influences fluidization and mixing, with broader distributions generally exhibiting lower minimum fluidization velocities and different mixing characteristics than monodisperse systems \cite{feng2017influence}. Binary mixtures provide a controlled framework for examining these effects while retaining the essential physics of size disparity.
The choice of fluidizing medium further influences particle mixing and segregation through differences in fluid--particle interactions.
The resulting hydrodynamics are inherently complex due to four-way coupling between the fluid and particulate phases, involving fluid forces, particle feedback on the fluid, and inter-particle collisions and contacts.
Understanding how particle size and fluidizing medium jointly influence mixing and hydrodynamics is therefore important for equipment design and operation.

Over the past few decades, numerous experimental and numerical investigations have examined the characteristics of gas--solid and liquid--solid fluidized beds \cite{xie2021cfd, luo2015cfd, wang2024cfd, islam2021effect, ma2017cfd}. 
Building on this foundation, mixing in binary beds have also been explored using either air or water as the fluidizing medium \cite{xie2021cfd, hoorijani2024comparative}. These studies collectively indicate that fluid properties strongly influence hydrodynamics, drag forces and particle velocities \cite{lu2015flow}. 

However, the emphasis in much of this literature has remained on such \emph{global} performance indicators (e.g., minimum fluidization velocity, bed expansion and pressure drop), which do not uniquely reveal the underlying mixing and segregation mechanisms in heterogeneous mixtures. 
Consequently, systematic, medium-to-medium comparisons that interrogate \emph{local} meso-scale and particle-scale flow dynamics---such as specie velocity fields, volume fraction, velocity fluctuations, local mixing rates, and microstructural organization (e.g., clustering or radial distribution characteristics)---in binary systems are still relatively scarce.
Addressing this gap is important because the fluidizing medium directly controls interphase momentum transfer and dissipation, thereby shaping the local mixing mechanisms that ultimately determine reactor performance.

Numerical simulation provides insight into granular flow dynamics and particle-scale interactions that are difficult to characterize experimentally. The two-fluid model (TFM) \cite{islam2021bed, puhan2021cfd} and computational fluid dynamics coupled with the discrete element method (CFD--DEM) \cite{liu2016cfd, moliner2019cfd} are widely used approaches. TFM treats both phases as interpenetrating continua, offering computational efficiency at the expense of particle-scale resolution. In contrast, CFD--DEM models the fluid as a continuum while tracking individual particles, resolving their motion, collisions, and fluid interactions, and is therefore well suited to particle-scale studies of fluidization.

The present study employs CFD--DEM to investigate the influence of fluidizing media on binary fluidized beds. 
A three-dimensional cylindrical column containing equal-density particles of two distinct diameters is simulated using air and water as the fluidizing media. 
Rather than focusing solely on global measures such as minimum fluidization velocity or bed expansion, the study links macroscopic bed response (bed height and pressure drop) to particle-scale dynamics, including solids distribution, particle motion and circulation, velocity fluctuations, microstructure, mixing kinetics, and particle trajectories.
This systematic comparison of gas--solid and liquid--solid fluidization provides insight into the role of fluid properties in governing mixing and flow behavior.

The remainder of this paper is organized as follows. Section~\ref{sec2} presents the mathematical formulation and CFD--DEM framework, followed by model validation in Section~\ref{sec3p1}. Section~\ref{sec4p} describes the computational setup, and Section~\ref{sec5} presents and discusses the results. Finally, Section~\ref{sec6} summarizes the main conclusions, limitations, and directions for future work.

\section{Methodology}
\label{sec2}
This section presents the CFD--DEM formulation used to model the fluidized bed. The fluid phase is solved on an Eulerian mesh, while individual particles are tracked in a Lagrangian framework, with the phases coupled through local volume averaging and interphase momentum exchange.

\subsection{Fluid phase}\label{sec2p1} 

The fluid phase is governed by the volume-averaged incompressible Navier--Stokes equations:
\begin{equation}
\frac{\partial (\alpha_f\rho_f )}{\partial t} + \nabla \cdot (\alpha_f \rho_f \mathbf{u}_f)=0, \label{eq1a}
\end{equation}
\begin{equation}
\frac{\partial (\alpha_f \rho_f \mathbf{u}_f)}{\partial t} + \nabla \cdot (\alpha_f \rho_f \mathbf{u}_f \mathbf{u}_f)= -\alpha_f \nabla p 
- \mathbf{K}_{sl} + \nabla \cdot(\alpha_f \bm{\tau}_f).
\label{eq2}
\end{equation}
Here, {$\alpha_f$} is the fluid volume fraction, $\rho_f$ is the fluid density, {$\mathbf{u}_f$} is the fluid velocity, and $p$ is the fluid pressure field, with the hydrostatic contribution excluded (i.e. the gravity term absorbed into modified pressure). 
The interphase momentum exchange term is:
\begin{equation}
\mathbf{K}_{sl} = \frac{ \sum_{i=1}^{N} \mathbf{F}_{f \rightarrow p}}{V_c}, \label{3}
\end{equation}
where {${V_c}$} is the computational-cell volume and $N$ the number of particles within the cell. For a Newtonian fluid, the deviatoric stress tensor is:
\begin{equation}
\bm{\tau}_f = \mu_f \left( \nabla \mathbf{u}_f + \nabla \mathbf{u}_f^{\mathrm{T}} \right)
- \frac{2}{3} \mu_f \left( \nabla \!\cdot\! \mathbf{u}_f \right) \mathbb{I},
\label{4}
\end{equation}
where $\mu_f$ is the dynamic viscosity and $\mathbb{I}$ is the identity tensor.

\subsection{Solid phase}\label{sec2p2}
The solid phase is resolved using DEM, with particle translation and rotation governed by the Newton--Euler equations:
\begin{equation}
m_i \frac{d \mathbf{u}_i}{d t}
= \sum \mathbf{F}_{c}^{\,i} +  \mathbf{F}_{f \rightarrow p} + m_i \mathbf{g},
\label{eq:dem_trans}
\end{equation}
\begin{equation}
\mathbf{I}_i \frac{d \bm{\omega}_i}{d t}
= \sum \mathbf{T}_{c}^{\,i},
\label{eq:dem_rot}
\end{equation}
where $m_i$, $\mathbf{u}_i$, $\bm{\omega}_i$, and $\mathbf{I}_i$ denote the mass, translational velocity, angular velocity, and moment of inertia tensor of particle $i$, respectively.
The term $\sum \mathbf{F}_{c}^{\,i}$ denotes the total contact force on particle $i$ arising from particle--particle and particle--wall collisions (normal and tangential components), and $\sum \mathbf{T}_{c}^{\,i}$ is the total contact torque acting on particle $i$, including the torque generated by tangential contact forces and rolling resistance during particle--particle and particle--wall interactions.

The fluid-to-particle force comprises drag, pressure-gradient, and buoyancy contributions:
\begin{equation}
\mathbf{F}_{f \rightarrow p}^{\,i}
=
\mathbf{F}_{d}^{\,i}
+
\mathbf{F}_{\nabla p}^{\,i}
+
\mathbf{F}_{b}^{\,i},
\label{eq:fluid_particle_force}
\end{equation}
The pressure-gradient force is obtained from the modified pressure field, while Archimedean buoyancy is accounted for separately.

\subsection{Fluid--solid interaction}\label{sec2p3}
The drag force is modeled using the Di Felice correlation \cite{difelice1994voidage}:

\begin{equation}
\mathbf{F}_{d}^{\,i}
= \frac{1}{2}\, C_d \left(\frac{\pi d_i^2}{4}\right)\rho_f\,\alpha_f^{\,2-\beta}\,|\mathbf{U}_r|\,\mathbf{U}_r
= \frac{\pi}{8}\,C_d\,\rho_f\,d_i^2\,\alpha_f^{\,2-\beta}\,|\mathbf{U}_r|\,\mathbf{U}_r,
\label{eq:drag_def}
\end{equation}
where $d_i$ is the particle diameter and the slip velocity is:
\begin{equation}
\mathbf{U}_r = \mathbf{u}_f - \mathbf{u}_i,
\label{eq:slip_vel}
\end{equation}
with $\mathbf{u}_f$ interpolated at the particle location.

The Di Felice correlation introduces the exponent $\beta$ as a function of particle Reynolds number, accounting for hindered settling and porosity effects in dense suspensions:
\begin{equation}
\beta = 3.7 - 0.65 \exp\!\left[-\frac{\left(1.5-\log \mathrm{Re}_p\right)^2}{2}\right],
\label{eq:beta_difelice}
\end{equation}
where the particle Reynolds number is computed individually for each particle using its corresponding particle diameter, $d_i$, as:
\begin{equation}
\mathrm{Re}_{p}
=
\frac{\rho_f d_i |\mathbf{U}_r|}{\mu_f}.
\label{eq:rep_def}
\end{equation}
The Di Felice drag coefficient is then evaluated as:
\begin{equation}
C_d =
\left(
0.63+\frac{4.8}{\sqrt{\mathrm{Re}_p}}
\right)^2.
\label{eq:cd_difelice}
\end{equation}

For the present bidisperse system, $\mathrm{Re}_p$, $C_d$, and the drag force are evaluated using the diameter of each particle.
The particle--particle and particle-–wall interactions were modeled using the Hertz–Mindlin contact model, which accounts for nonlinear elastic normal contact forces and tangential frictional forces. The fluid phase was treated as laminar for both the air- and water-fluidized bed simulations.

\subsection{Algorithm}\label{sec2p4}
\begin{figure*}[!htbp]
	\centering
	\includegraphics[width=0.9\textwidth]{
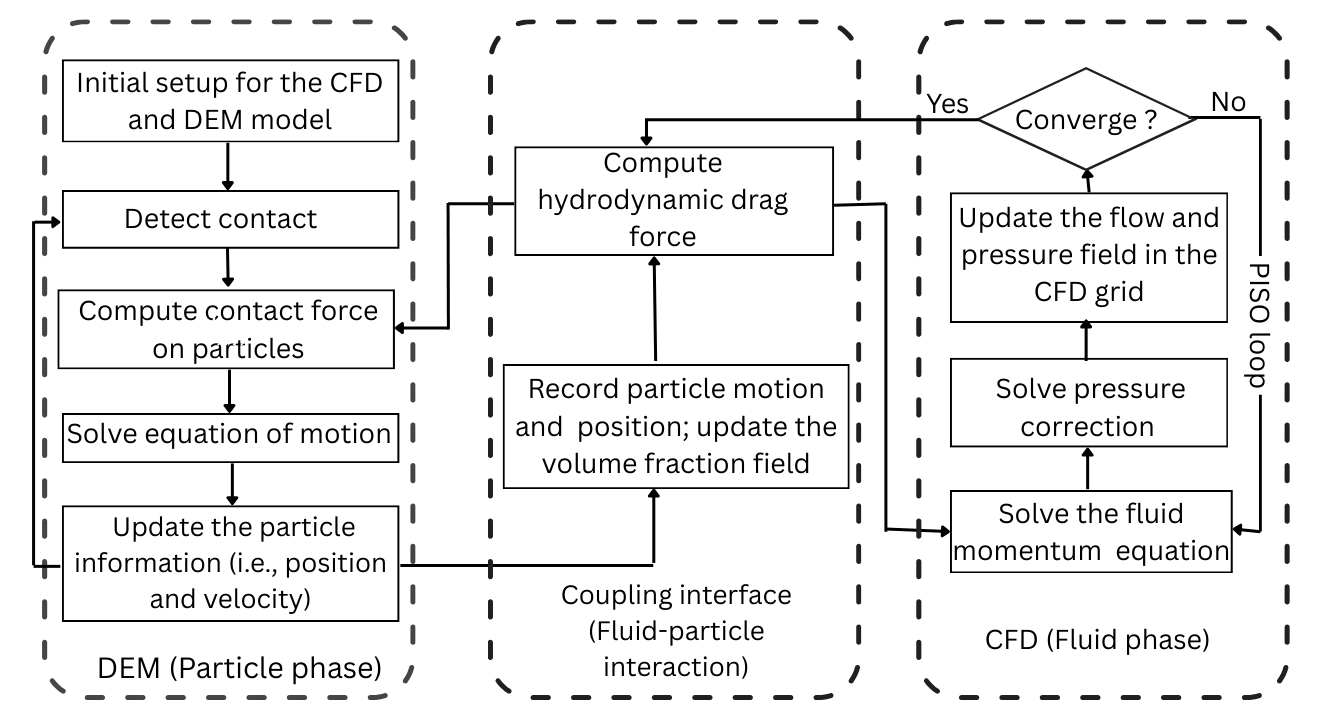}
	\caption{Schematic representation of the CFD--DEM coupling algorithm.}
	\label{fig:A}
\end{figure*}
The CFD and DEM solvers are coupled sequentially at prescribed intervals (Figure~\ref{fig:A}).
The DEM timestep $\Delta t_{\mathrm{DEM}}$ is smaller than the CFD timestep $\Delta t_{\mathrm{CFD}}$, with:
\begin{equation}
\Delta t_{\mathrm{CFD}}=N_c\Delta t_{\mathrm{DEM}},
\end{equation}
where $N_c$ is the number of DEM substeps per CFD update.

At each coupling step, particle positions are mapped to the CFD mesh to determine the local fluid volume fraction and interphase momentum exchange. The fluid equations are then advanced using the PISO algorithm, and the updated fluid fields and hydrodynamic forces are transferred to the particles. The DEM solver subsequently advances particle motion for $N_c$ substeps, updating contact forces and torques at each substep, before the coupling cycle is repeated.

The simulations were performed using CFDEM\textsuperscript{\textregistered} 3.8.1, coupling OpenFOAM 5.x with LIGGGHTS\textsuperscript{\textregistered} 3.8.0. PISO employed two correctors, with PCG/DIC and PBiCG/DILU tolerances of $10^{-6}$ and $10^{-5}$ for pressure and velocity, respectively.

\section{Model validation}
\label{sec3p1}

The CFD--DEM framework was assessed for both liquid--solid and gas--solid fluidization. For the liquid--solid case, the model was validated against the experiments of \citet{khan2016pressure} for a bidisperse water-fluidized bed. For gas--solid fluidization, experimental data matching the adopted configuration were unavailable; therefore, the model was benchmarked against the published numerical results of \citet{sakai2014verification}. The latter comparison is intended to verify consistency with an established CFD--DEM solution rather than constitute independent experimental validation.

\begin{figure}[!htbp]
    \centering
    \includegraphics[width=0.8\columnwidth]{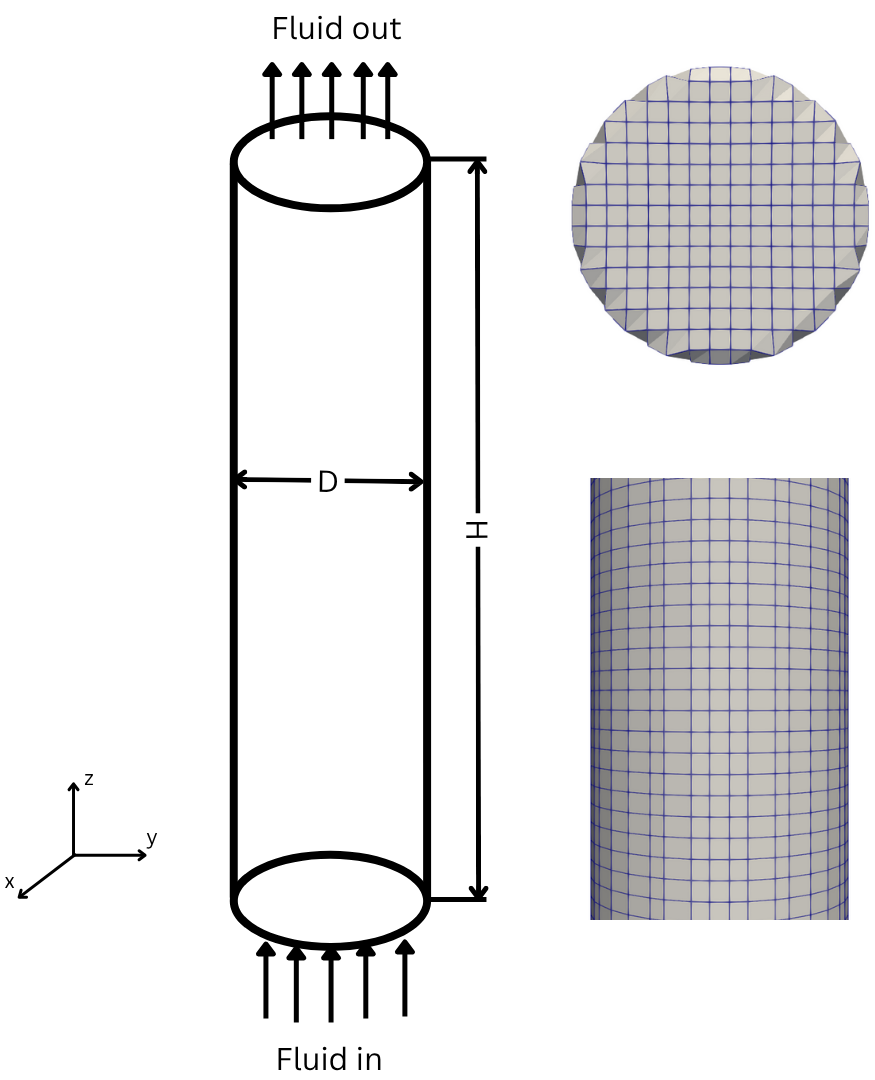}
    \caption{Geometry and mesh used for liquid--solid validation and for the simulations in this work.}
    \label{fig:1}
\end{figure}

For the liquid--solid validation, the cylindrical column of \citet{khan2016pressure}, with height $H=0.7$~m and diameter $D=0.05$~m, was reproduced (Figure~\ref{fig:1}). The bed contained equal masses (120~g each) of 3 and 8~mm spherical particles with density $\rho_p=\SI{2230}{\kilo\gram\per\cubic\meter}$, initially arranged with the smaller particles above the larger particles. Water was introduced uniformly from the base at the prescribed superficial velocity. The computational mesh for validation was selected based on a mesh-independence study, with an approximate cell size of $0.003~\text{m} \times 0.003~\text{m} \times 0.003~\text{m}$.

\begin{figure*}[!htbp]  
	\centering
	\includegraphics[width=1\textwidth]{
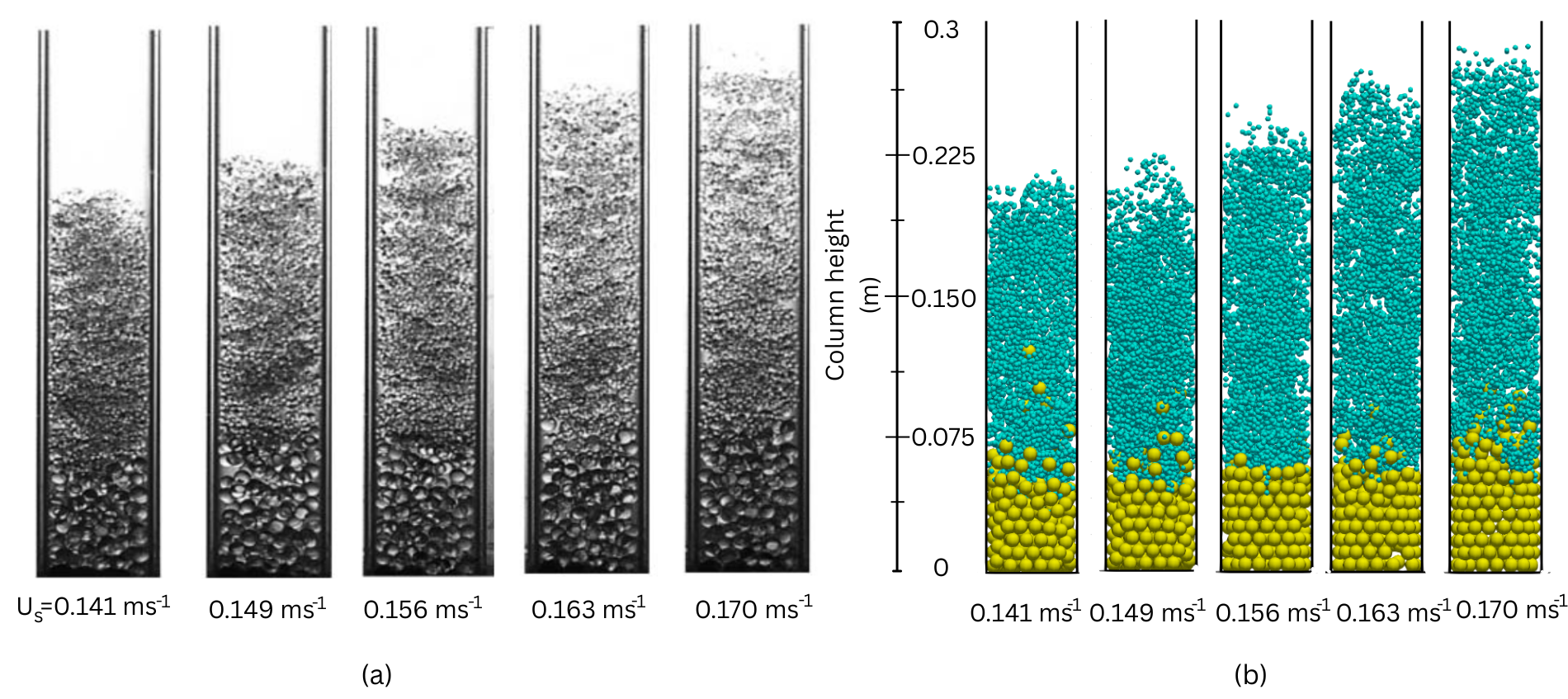}
	\caption{Qualitative comparison of the binary liquid--solid fluidized bed at increasing superficial velocities: (a) experiments of \citet{khan2016pressure}; (b) corresponding CFD--DEM simulations.}
	\label{fig:3}
\end{figure*}

\begin{figure}[!htbp]
	\centering
	\includegraphics{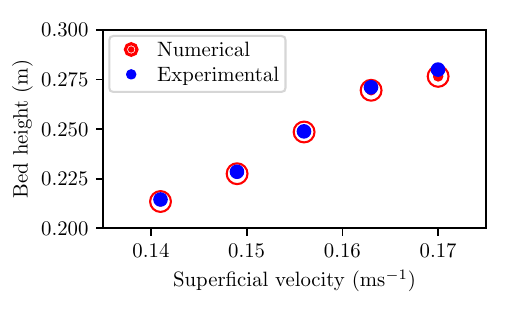}
	\caption{Comparison of simulated and experimental \cite{khan2016pressure} bed heights at different superficial velocities.}
	\label{fig:4}
\end{figure}

Validation was performed at $U_s=0.141$, 0.149, 0.156, 0.163, and \SI{0.170}{\meter\per\second}, with each case simulated for 20~s to obtain a statistically steady expanded bed. The simulations reproduce the experimentally observed bed expansion and particle redistribution over the range of superficial velocities (Figure~\ref{fig:3}). Quantitatively, the predicted bed heights agree closely with the measurements and capture their increase with superficial velocity (Figure~\ref{fig:4}), supporting the suitability of the adopted fluid--particle coupling and drag formulation for liquid--solid fluidization.

For the gas--solid benchmark, the configuration of \citet{sakai2014verification} was reproduced using a domain of $20\times5\times50$~mm$^3$ with a uniform 1~mm CFD grid. The bed contained 540,000 spherical particles of diameter 150~$\mu$m and density \SI{2500}{\kilo\gram\per\cubic\meter} packed randomly. The gas density and dynamic viscosity were \SI{1.0}{\kilo\gram\per\cubic\meter} and $1.8\times10^{-5}$~Pa,s, respectively, and the superficial gas velocity was 0.07~m\,s$^{-1}$.

\begin{figure}[!htbp]
\centering
\includegraphics{
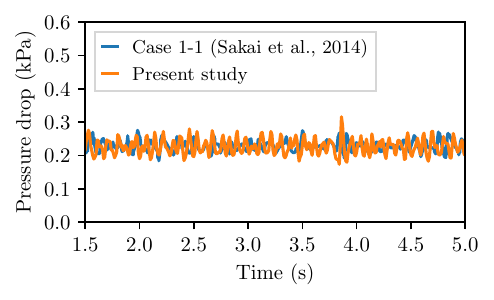}
\caption{Comparison of the present pressure-drop prediction with the numerical benchmark of \citet{sakai2014verification}.}
\label{fig:gas_vali}
\end{figure}

The predicted pressure-drop evolution agrees closely with the published numerical results (Figure~\ref{fig:gas_vali}), demonstrating that the present implementation reproduces the established bubbling gas--solid benchmark. Together with the experimental liquid--solid validation, this provides a basis for applying the CFD--DEM framework to the comparative simulations considered here.

\section{Computational setting}\label{sec4p}

This section describes the numerical configuration employed for the new CFD--DEM simulations, including the computational geometry and mesh, initial and boundary conditions, fluid properties, numerical parameters, and grid independence assessment.

\subsection{Model geometry and mesh}\label{sec4p1}

The simulations employed a cylindrical column of diameter $D=0.14$~m and height $H=0.45$~m, with the bottom and top boundaries serving as the fluid inlet and outlet, respectively (Figure~\ref{fig:1}). The predominantly hexahedral mesh was generated using \texttt{blockMesh} and \texttt{snappyHexMesh}.

\subsection{Initial conditions}\label{sec4p2}

The bed contained equal masses (1~kg each) of 3 and 4~mm spherical particles with density $\rho_p=\SI{2230}{\kilo\gram\per\cubic\meter}$. The larger particles were initially placed below the smaller particles and allowed to settle under gravity without fluid flow. The resulting mechanically stable configuration was used to initialize all fluidization simulations. Particles were assumed cohesionless, and the initial fluid velocity was zero.

\subsection{Boundary conditions}\label{sec4p3}

\begin{table}[htbp]
\centering
\caption{Boundary conditions for velocity and pressure fields}
\renewcommand{\arraystretch}{1.5}
\setlength{\tabcolsep}{10pt}     
\begin{tabular}{lccc}
\hline
\textbf{Property} & \textbf{Inlet} & \textbf{Wall} & \textbf{Outlet} \\
\hline
\textbf{Velocity, $\mathbf{u}_f$}
    & $U_s \hat{k}$ 
    & No-slip 
   & $\frac{\partial \mathbf{u}_f}{\partial n} = 0$ \\
\textbf{Pressure, $p$} 
    & $\frac{\partial p}{\partial n} = 0$ 
    & $\frac{\partial p}{\partial n} = 0$ 
    & $p = p_0$ \\
\hline
\end{tabular}
\label{table1}
\end{table}
Boundary conditions for the fluid phase are summarized in Table~\ref{table1}. A uniform superficial velocity $U_s$ was prescribed at the inlet, while the outlet pressure was fixed at $p_0$. The cylindrical wall was treated as no-slip, and all boundaries were rigid and impenetrable to particles.
For the DEM simulations, all boundaries were treated as rigid, impenetrable walls for particle motion.

\subsection{Numerical parameters}\label{sec4p5a}

The CFD timestep was selected to maintain the Courant number below 0.5:
\begin{equation}
C_o = \frac{|\mathbf{u}_f| \Delta t_{\mathrm{CFD}}}{\Delta x},
\label{eq:courant}
\end{equation}
while the DEM timestep was constrained by the Rayleigh criterion \cite{thornton2015granular},
\begin{equation}
\Delta t_{\mathrm{max}} = \frac{\pi r_{\mathrm{min}}}{0.163\nu + 0.8766}\sqrt{\frac{\rho_p}{G}},
\label{eq:rayleigh}
\end{equation}
where $G$ is the shear modulus.
A DEM timestep of $5\times10^{-6}$~s and coupling interval $N_c=40$ were used (i.e. information was exchanged every 40 DEM time steps.), giving $\Delta t_{\mathrm{CFD}}=2\times10^{-4}$~s. The principal physical and numerical parameters are summarized in Table~\ref{table2}.

\begin{table}[htbp]
\centering
\caption{CFD-DEM simulation parameters}
\begin{tabular}{lcc}
\hline
\textbf{Parameter} & \textbf{Value} & \textbf{Unit} \\
\hline
\multicolumn{3}{l}{\textbf{Cell geometry}} \\
Column height, $H$         & $0.45$ &  m \\
Column diameter, $D$     & $0.14$ & m \\
\multicolumn{3}{l}{\textbf{Solid phase}} \\
Small particle diameter, $d_1$ & $3$ &  mm \\
Large particle diameter, $d_2$ & $4$ &  mm \\
Particles density, $\rho_p$     & $2230$ & \si{\kilo\gram\per\cubic\meter} \\
Young's modulus, $E$         & $1.0 \times 10^7$ &  Pa \\
Poisson's ratio, $\nu$       & $0.3$ & -\\
Restitution coefficient, $e$ & $0.6$ & - \\
Sliding friction coefficient, $\mu_s$ & $0.3$ & - \\
Rolling friction coefficient, $\mu_r$ & $0.002$ & - \\
Initial bed height, $H_i$ & $0.093$ & m \\
Initial solid holdup, $\alpha_s$ & 0.626 & -\\
\multicolumn{3}{l}{\textbf{Fluid phase}} \\
Liquid density, $\rho_w$ & $1000$ & \si{\kilo\gram\per\cubic\meter} \\
Liquid viscosity, $\mu_w$ & $0.001$ & $\text{Pa$\cdot$s}$ \\
Gas density, $\rho_a$ & $1.2$ & \si{\kilo\gram\per\cubic\meter} \\
Gas viscosity, $\mu_a$ & $1.9 \times 10^{-5}$ & $\text{Pa$\cdot$s}$ \\
\multicolumn{3}{l}{\textbf{Simulation setup}} \\
CFD time step, $\Delta t_{\text{CFD}}$ & $2.0 \times 10^{-4}$ & s \\
DEM time step, $\Delta t_{\text{DEM}}$ & $5.0 \times 10^{-6}$ & s \\
Simulation time & $20$ & s \\
\hline
\end{tabular}
\label{table2}
\end{table}

\subsection{Mesh sensitivity study}\label{sec4p6}

A mesh sensitivity study was performed to evaluate the influence of CFD mesh resolution on the predicted bed expansion height for both water- and air-fluidized beds. Simulations were conducted at a representative superficial velocity of $U_s = 3U_{mf}$ using three mesh resolutions: coarse ($31 \times 31 \times 84$), medium ($37 \times 39 \times 105$), and fine ($45 \times 47 \times 126$), where $U_{mf}$ denotes the minimum fluidization velocity. 

\begin{figure}[t]
	\centering
	\subfigure[Water case]{
		\includegraphics[width=0.48\textwidth]{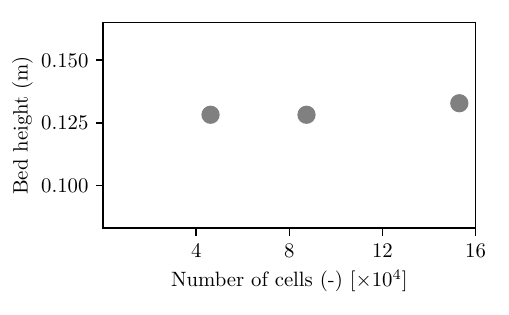}
		\label{fig:grid_water}
	}
	\hfill
	\subfigure[Air case]{
		\includegraphics[width=0.48\textwidth]{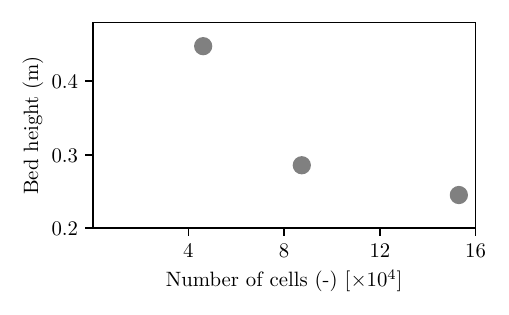}
		\label{fig:grid_air}
	}
    \caption{Bed expansion height versus number of cells at {$U_s = 3U_{mf}$}: (a) water and (b) air.}    	\label{fig:grid_convergence}

\end{figure}

Figure~\ref{fig:grid_convergence} shows the mesh-sensitivity assessment for the water- and air-fluidized beds at $U_s=3U_{mf}$. For water, the predicted bed height varies only marginally across the three meshes, indicating negligible mesh sensitivity. The air case exhibits greater sensitivity, particularly on the coarse mesh. However, the change in bed height decreases from approximately 35\% between the coarse and medium meshes to about 10\% between the medium and fine meshes, indicating convergence with mesh refinement. The medium mesh was therefore considered sufficiently resolved for the present comparative study and was adopted for both fluidizing media.

The selected mesh has an equivalent cell size of approximately 4.3~mm, giving $\Delta/d_p$ ratios of 1.43 and 1.08 for the 3 and 4~mm particles, respectively. Since the cell size is comparable to the particle diameter, the \texttt{divided} void-fraction model in CFDEM was employed to distribute particle volume among neighbouring cells, providing a smoother void-fraction field and more robust fluid--particle coupling \cite{peng2014influence,clarke2018investigation}.

The production simulations contained approximately 45,000 particles and were run on 48 CPU cores (Intel Xeon Cascade Lake). A typical 20~s simulation at the highest fluidization velocity required approximately 7~h of wall time.

\section{Results and discussion}\label{sec5}

This section systematically compares binary fluidization in liquid--solid (\emph{water}) and gas--solid (\emph{air}) systems under identical geometric and particle conditions. The comparison is made at matched normalized superficial velocities, $U_s/U_{mf}$, linking global bed response to the underlying local flow and particle-scale dynamics.

\subsection{Fluidization behavior}
\label{sec5p1}

Simulations were performed for water and air at six normalized superficial velocities, $U_s/U_{mf}=0.5$--$3.0$, for 20~s of physical time. The reference minimum fluidization velocity was estimated using the Wen--Yu correlation \cite{wen1966generalized}, which defines the particle $Re$ at incipient fluidization as:
\begin{equation}
Re_{mf} =
\left(A^2 + B\,\mathrm{Ar}\right)^{1/2} - A,
\label{eq:remf}
\end{equation}
with empirical constants $A=33.7$, $B=0.0408$, and the Archimedes number:
\begin{equation}
\mathrm{Ar} =
\frac{g d_p^3 \rho_f(\rho_p-\rho_f)}{\mu_f^2}.
\label{eq:archimedes}
\end{equation}
The corresponding minimum fluidization velocity is
\begin{equation}
U_{mf} =
\frac{\mu_f}{\rho_f d_p}
\left[
\left(A^2 + B\,\mathrm{Ar}\right)^{1/2} - A
\right].
\label{eq:umf}
\end{equation}

Using the smaller particle diameter as a reference, since fluidization starts when the smaller particles are lifted, gives $U_{mf}=0.029~\mathrm{m,s^{-1}}$ for water and $1.325~\mathrm{m,s^{-1}}$ for air. Since the Wen--Yu correlation is based on monodisperse beds, these values are used only to normalize the operating conditions; the effective $U_{mf}$ of the bidisperse bed is determined subsequently from the simulated pressure-drop response.
Also, the calculated values of $Re_{mf}$ for water- and air- systems were approximately 86 and 251, respectively, both of which are well within the applicability of the correlation ($Re_{mf} < 1000$).

\begin{figure*}[!htbp]
    \centering
    \subfigure[Water-fluidized bed]{
        \includegraphics[width=1.0\textwidth]{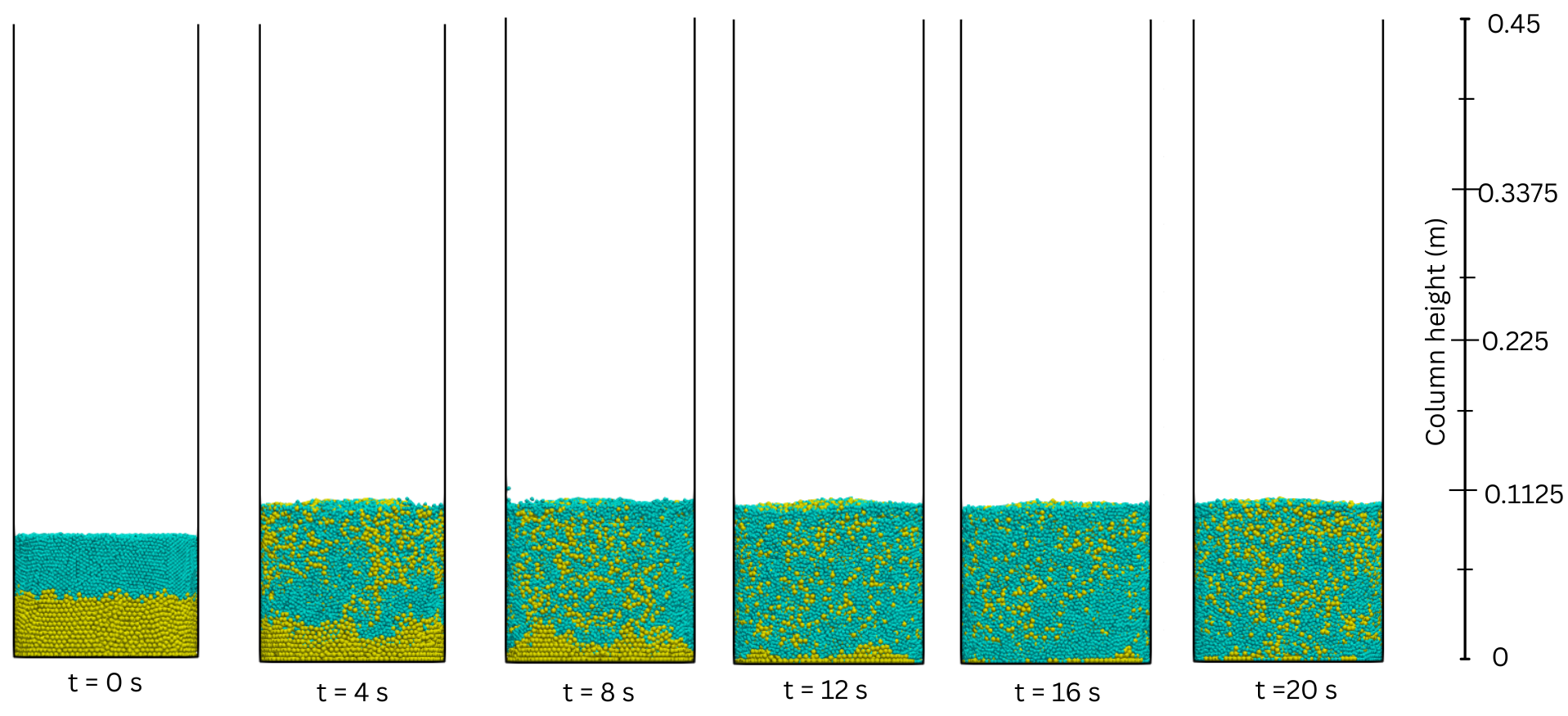}
        \label{fig:7}
    }

    \subfigure[Air-fluidized bed]{
        \includegraphics[width=1.0\textwidth]{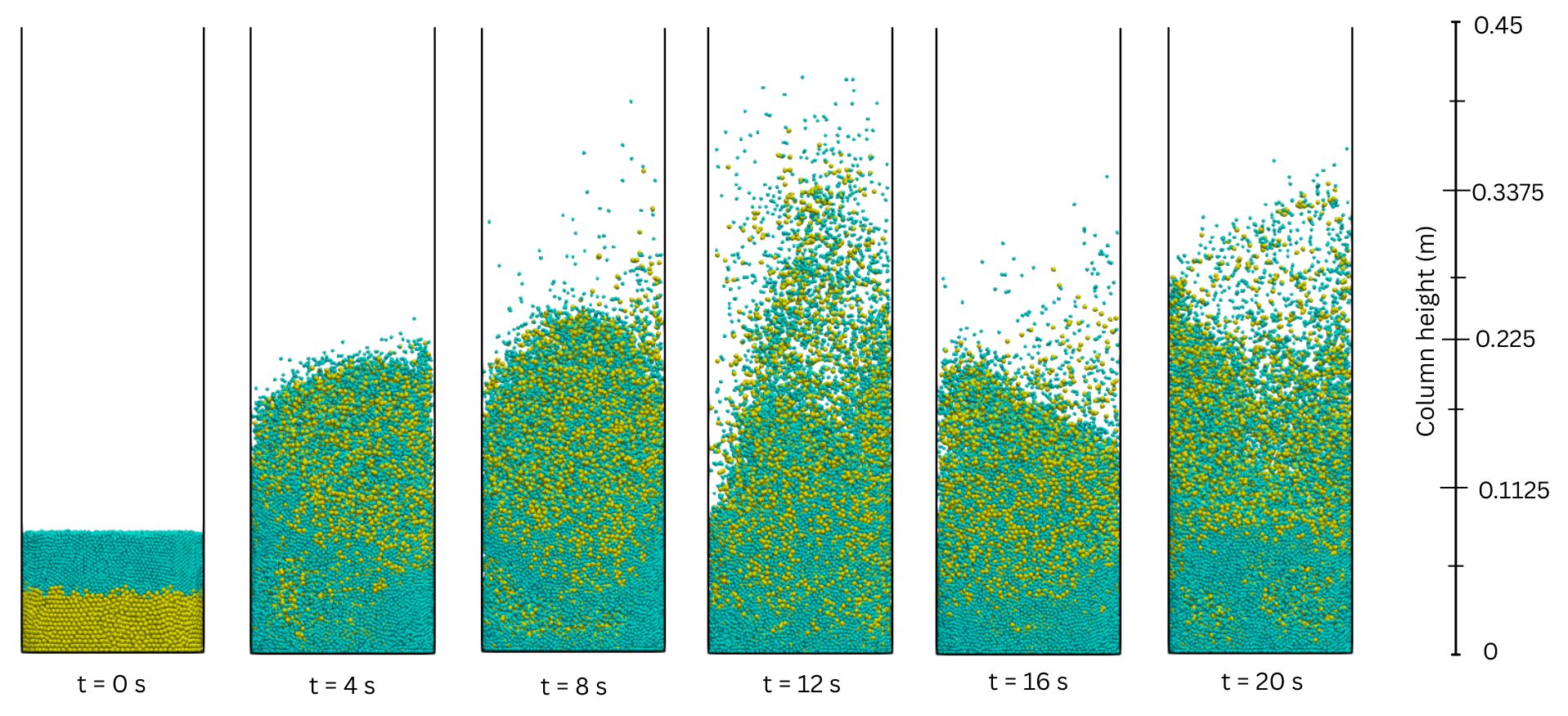}
        \label{fig:8}
    }

    \caption{Evolution of the binary fluidized bed at $U_s=3U_{mf}$ in
    (a) water and (b) air. The 3 and 4~mm particles are shown in green
    and yellow, respectively.}
    \label{fig:78}
\end{figure*}

Figures~\ref{fig:7} and~\ref{fig:8} compare the evolution of the water- and air-fluidized beds at $U_s=3U_{mf}$ from the same initially layered configuration. The water-fluidized bed expands gradually to a relatively stable, homogeneous state, with progressive mixing and a well-defined bed interface. 
The higher density and viscosity of water promote stronger interphase momentum transfer and viscous damping, suppressing fluctuations.

In contrast, the air-fluidized bed exhibits substantially greater expansion, large voids, and persistent particle rearrangement associated with bubble-like structures. The lower density and viscosity of air provide weaker hydrodynamic damping, resulting in greater particle mobility and pronounced spatial and temporal heterogeneity. Thus, even at the same $U_s/U_{mf}$, the two media produce distinctly different fluidization regimes: a relatively uniform and stable bed in water and a highly expanded and unsteady bed in air.

\subsection{Velocity field and fluctuations}
\label{sec5p2}

\begin{figure*}[!htbp]
	\centering
	\includegraphics[width=1.0\textwidth]{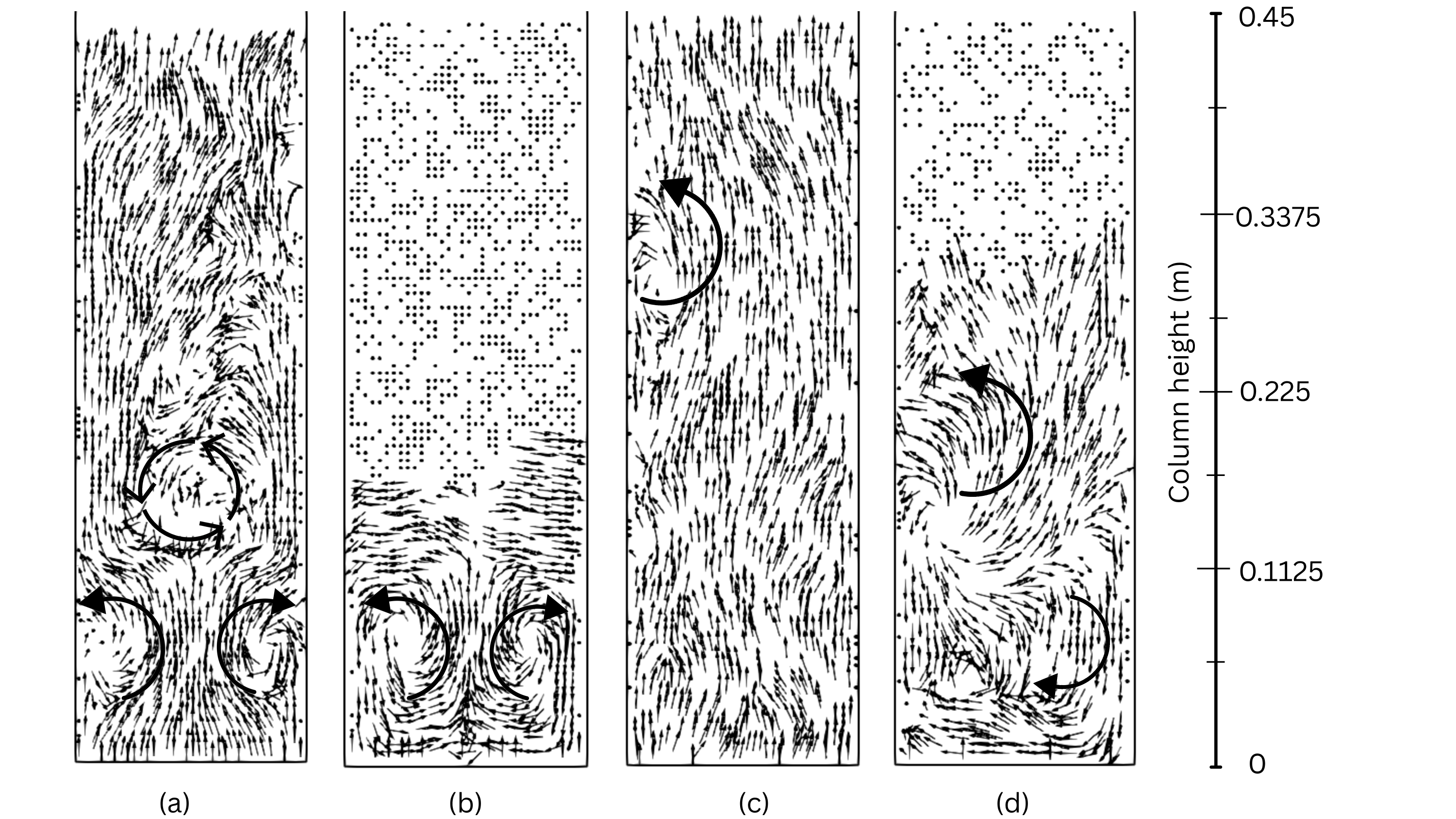}
	\caption{Velocity fields of the fluid and particle phases at $U_s=3U_{mf}$: (a,b) water and (c,d) air. Vectors show the projected in-plane velocity on the vertical mid-plane ($x = 0$) at $t$=17s.}
	\label{fig:vel_field}
\end{figure*} 
Figure~\ref{fig:vel_field} compares the instantaneous fluid and particle velocity fields at $U_s=3U_{mf}$. In the water-fluidized bed, the predominantly upward fluid motion is accompanied by localized recirculation in the dense lower region, with corresponding circulation cells evident in the particle field. This similarity indicates strong coupling between the fluid and particle motion.

In the air-fluidized bed, the gas velocity remains comparatively smooth and vertically aligned, whereas the particle field exhibits pronounced recirculation and greater spatial variability. The weaker correspondence between the two phases reflects the lower density and viscosity of air and the resulting weaker hydrodynamic coupling.

The flow unsteadiness was quantified using the root-mean-square axial fluid velocity fluctuation on the vertical mid-plane,
\begin{equation}
u_{f,z}' =
\frac{1}{N_c}
\sum_{c=1}^{N_c}
\sqrt{
\frac{1}{N_t}
\sum_{t=t_{bs}}^{t_{es}}
\left[
u_{f,z}(c,t)-\overline{u}_{f,z}(c)
\right]^2
},
\label{eq:fluid_velocity_fluctuation}
\end{equation}
where $N_c$ and $N_t$ are the numbers of sampled cells and time instances, respectively, and $c$ denotes the sampled cell. Statistics were evaluated over $t=5$--$20$~s at 1~s intervals.

\begin{figure*}[!htbp]
	\centering
	\subfigure[Water]{
		\includegraphics[width=0.48\textwidth]{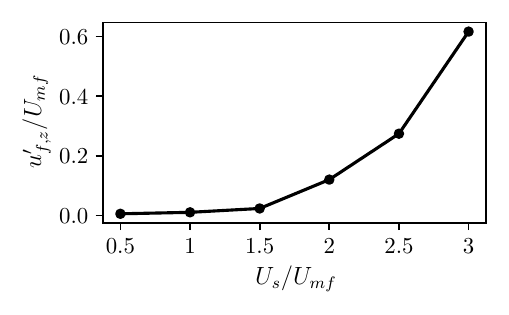}
		\label{fig:flu_water}
	}
	\hfill
	\subfigure[Air]{
		\includegraphics[width=0.48\textwidth]{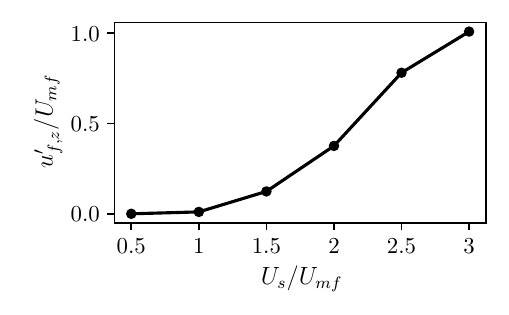}
		\label{fig:flu_air}
	}
	\caption{Variation of the dimensionless root-mean-square axial fluid velocity fluctuation, $u'_{f,z}/U_{mf}$, as a function of the normalized superficial velocity, $U_s/U_{mf}$ for (a) water and (b) air.}
	\label{fig:fluctuation}
\end{figure*}

Figure~\ref{fig:fluctuation} shows increasing velocity fluctuations with $U_s/U_{mf}$ in both media, with a pronounced rise above $U_s/U_{mf}=2$. The increase is substantially stronger in air, indicating a transition toward a more heterogeneous and dynamically active regime. The larger fluctuations are consistent with weaker hydrodynamic damping in air and the resulting stronger particle and flow disturbances \cite{peng2014forces, xin2025fluid}.

\begin{figure*}[!htbp]
  \centering

  \subfigure[ Water, $U_s/U_{mf}=2$]{
    \includegraphics[width=0.47\textwidth]{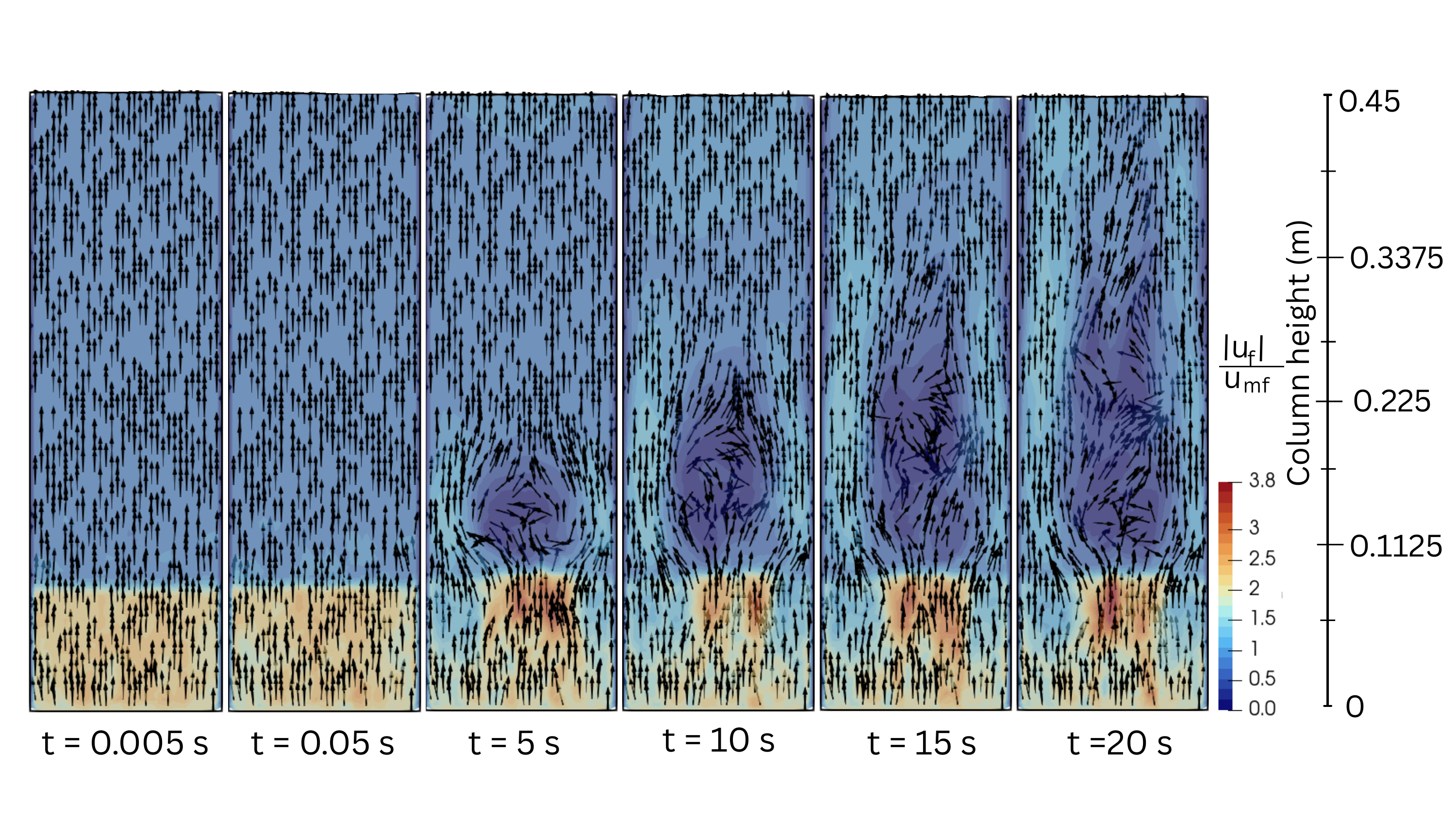}
    \label{fig:water_2umf1}
  }
  \hfill
  \subfigure[ Air, $U_s/U_{mf}=2$]{
    \includegraphics[width=0.47\textwidth]{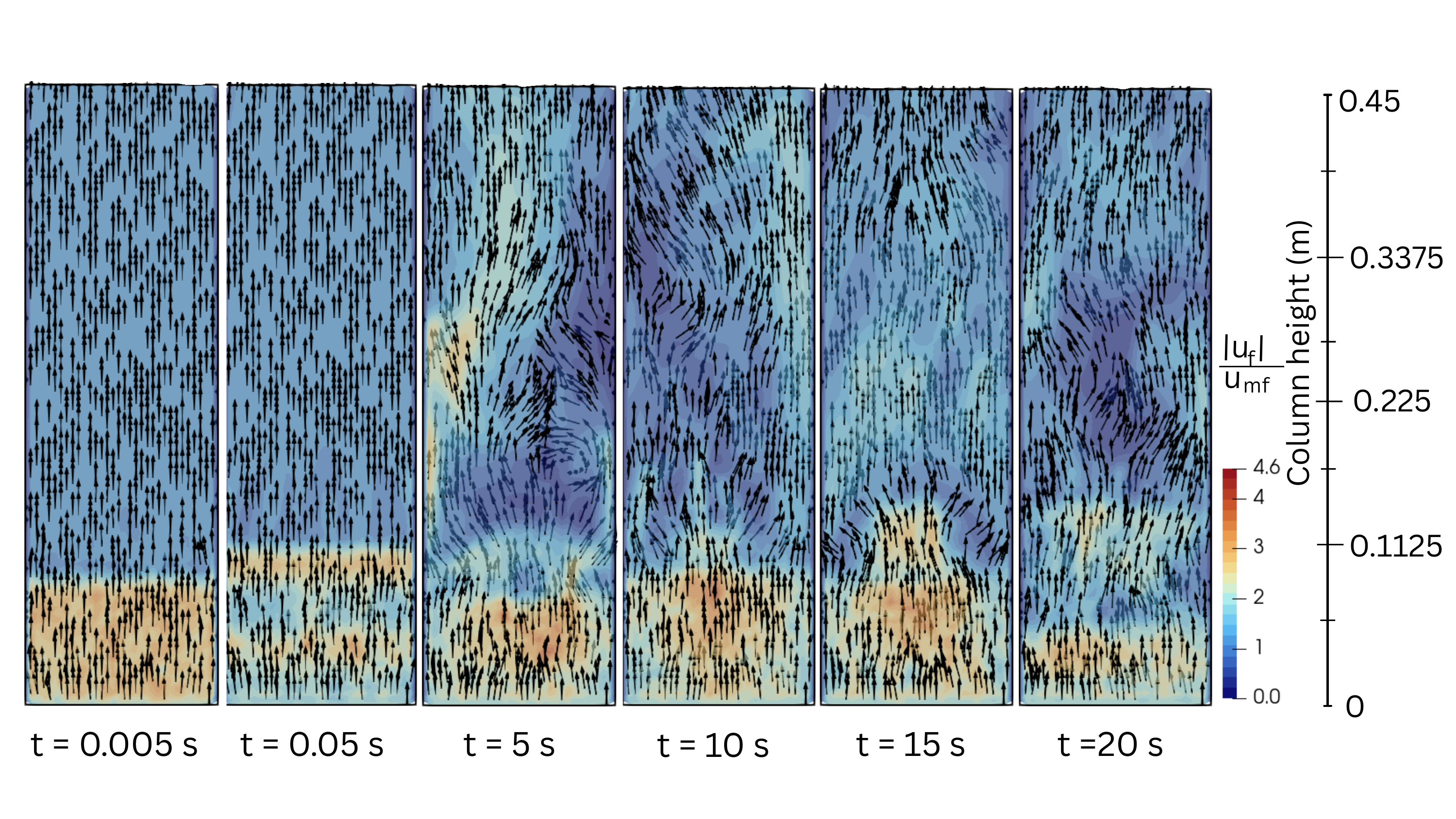}
    \label{fig:air_2umf1}
  }

  \subfigure[ Water, $U_s/U_{mf}=2.5$]{
    \includegraphics[width=0.47\textwidth]{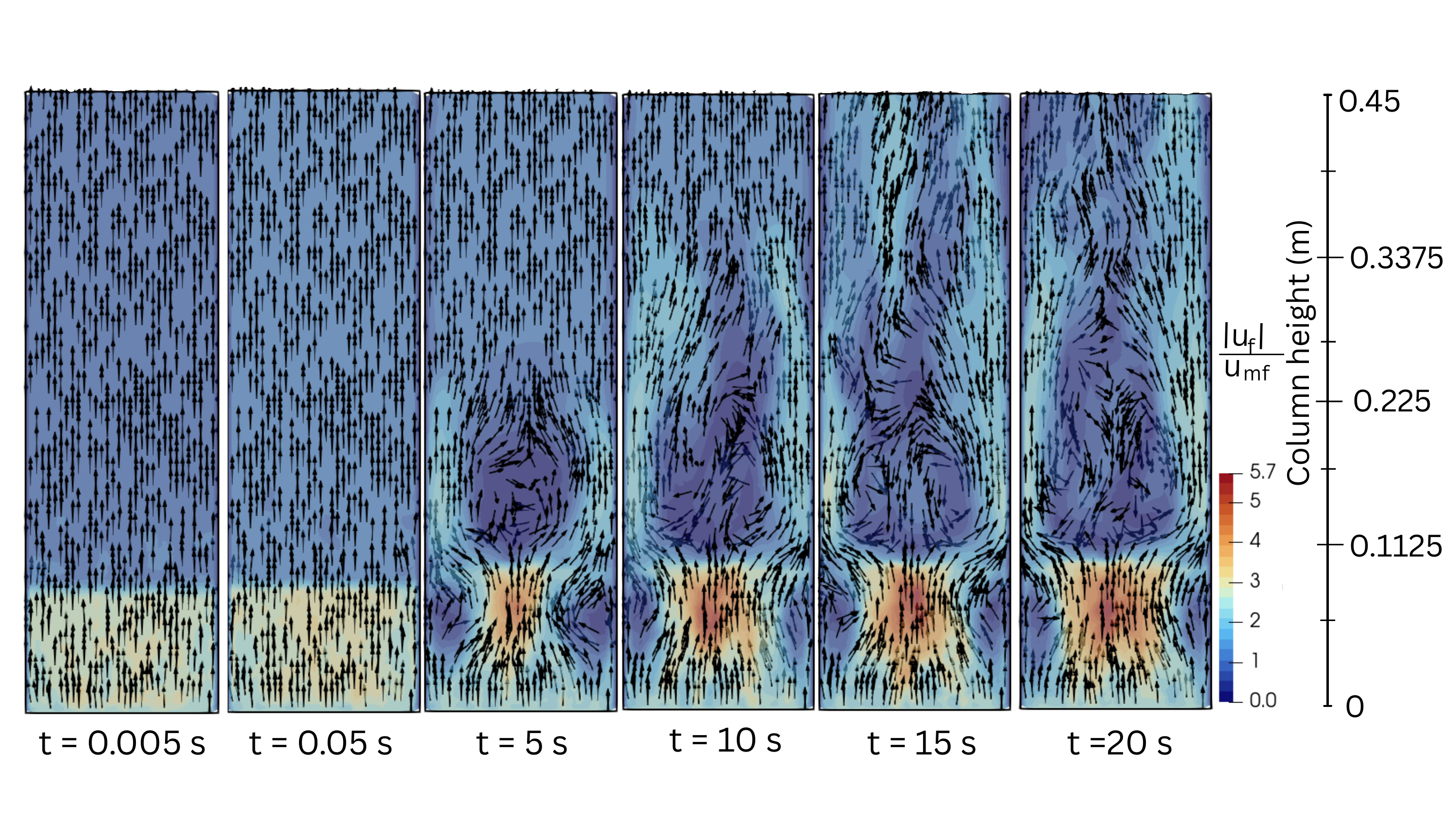}
    \label{fig:water_2.5umf1}
  }
  \hfill
  \subfigure[ Air, $U_s/U_{mf}=2.5$]{
    \includegraphics[width=0.47\textwidth]{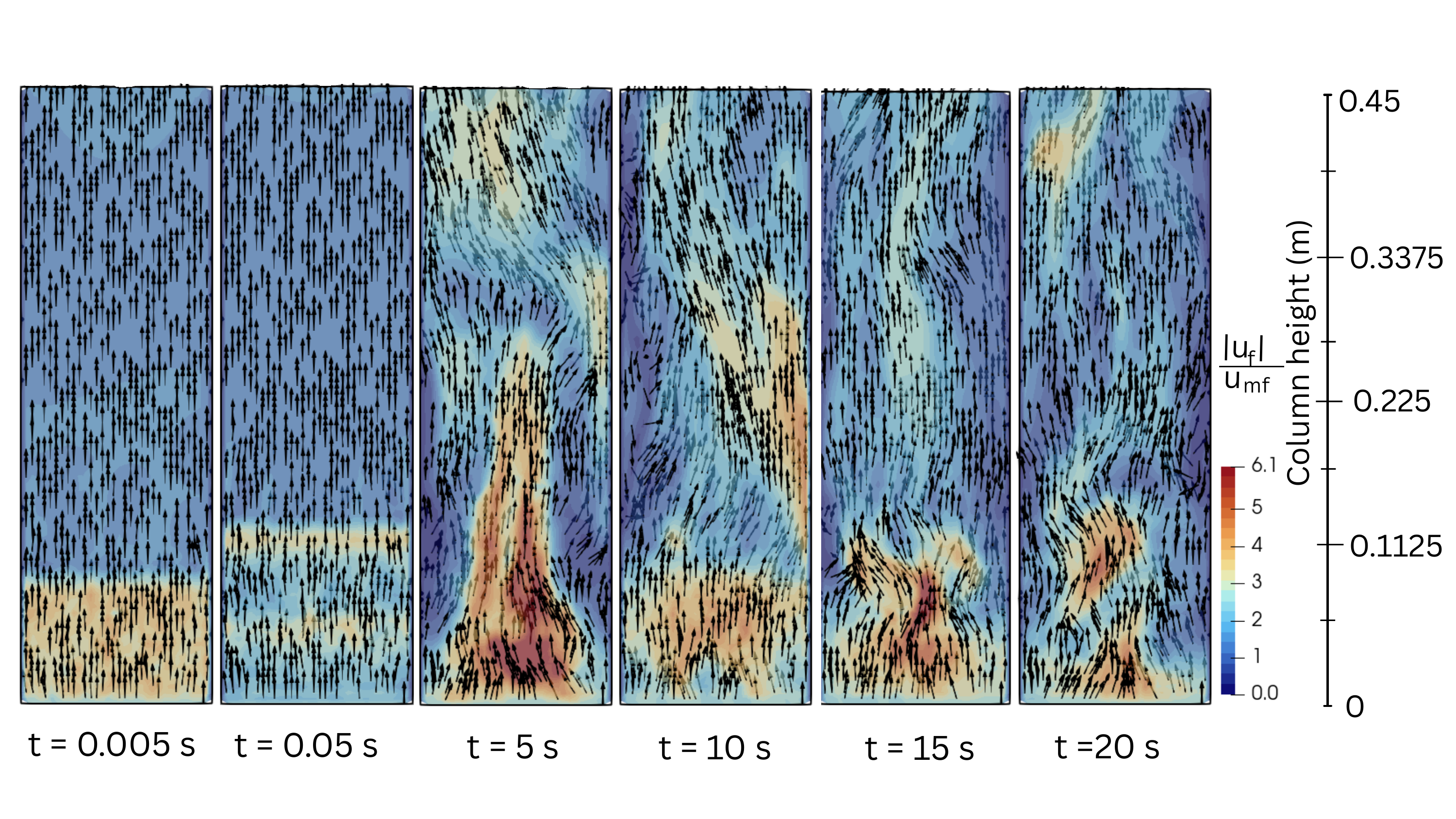}
    \label{fig:air_2.5umf1}
  }

  \subfigure[ Water, $U_s/U_{mf}=3$]{
    \includegraphics[width=0.47\textwidth]{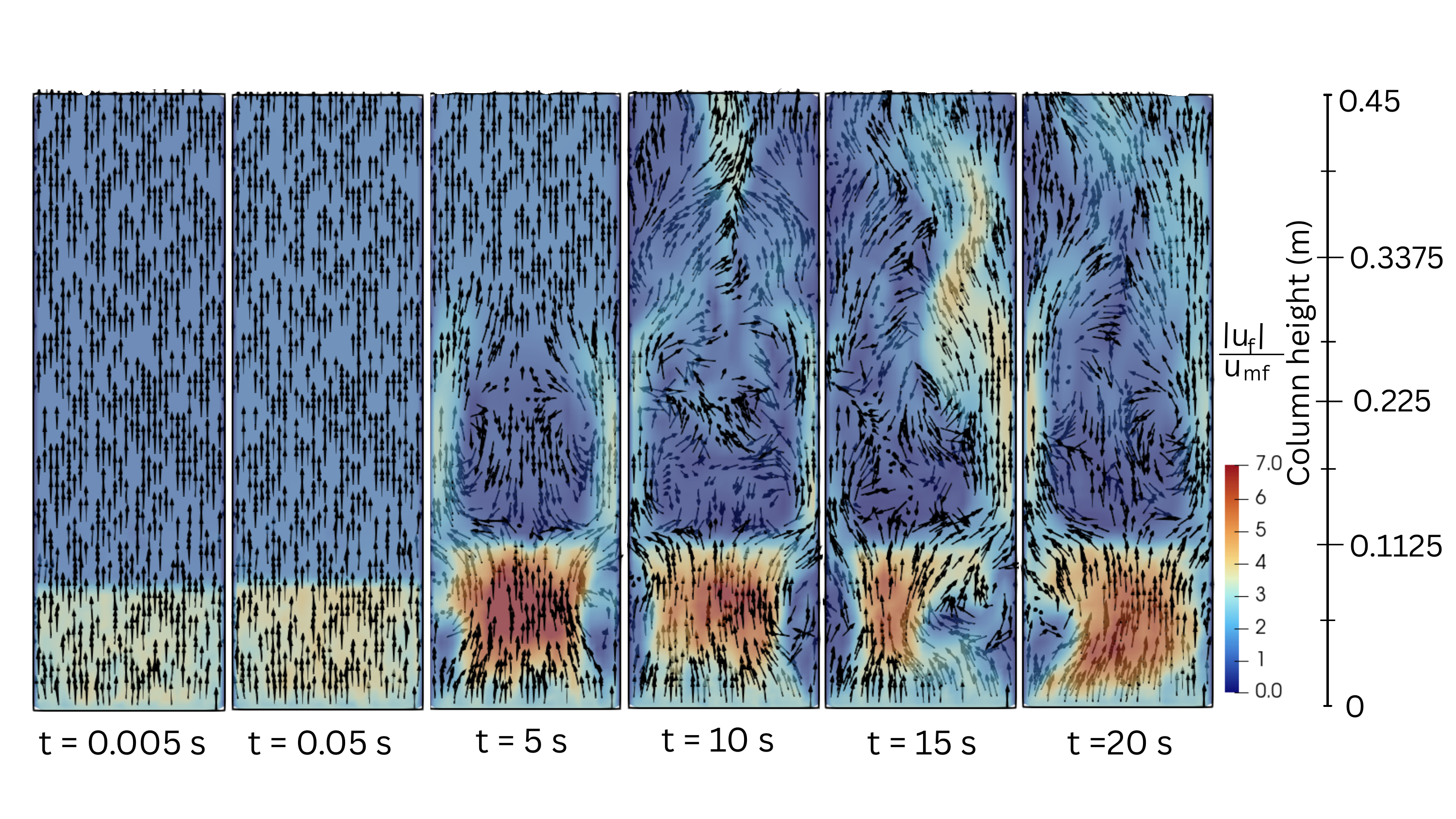}
    \label{fig:water_3umf1}
  }
  \hfill
  \subfigure[ Air, $U_s/U_{mf}=3$]{
    \includegraphics[width=0.47\textwidth]{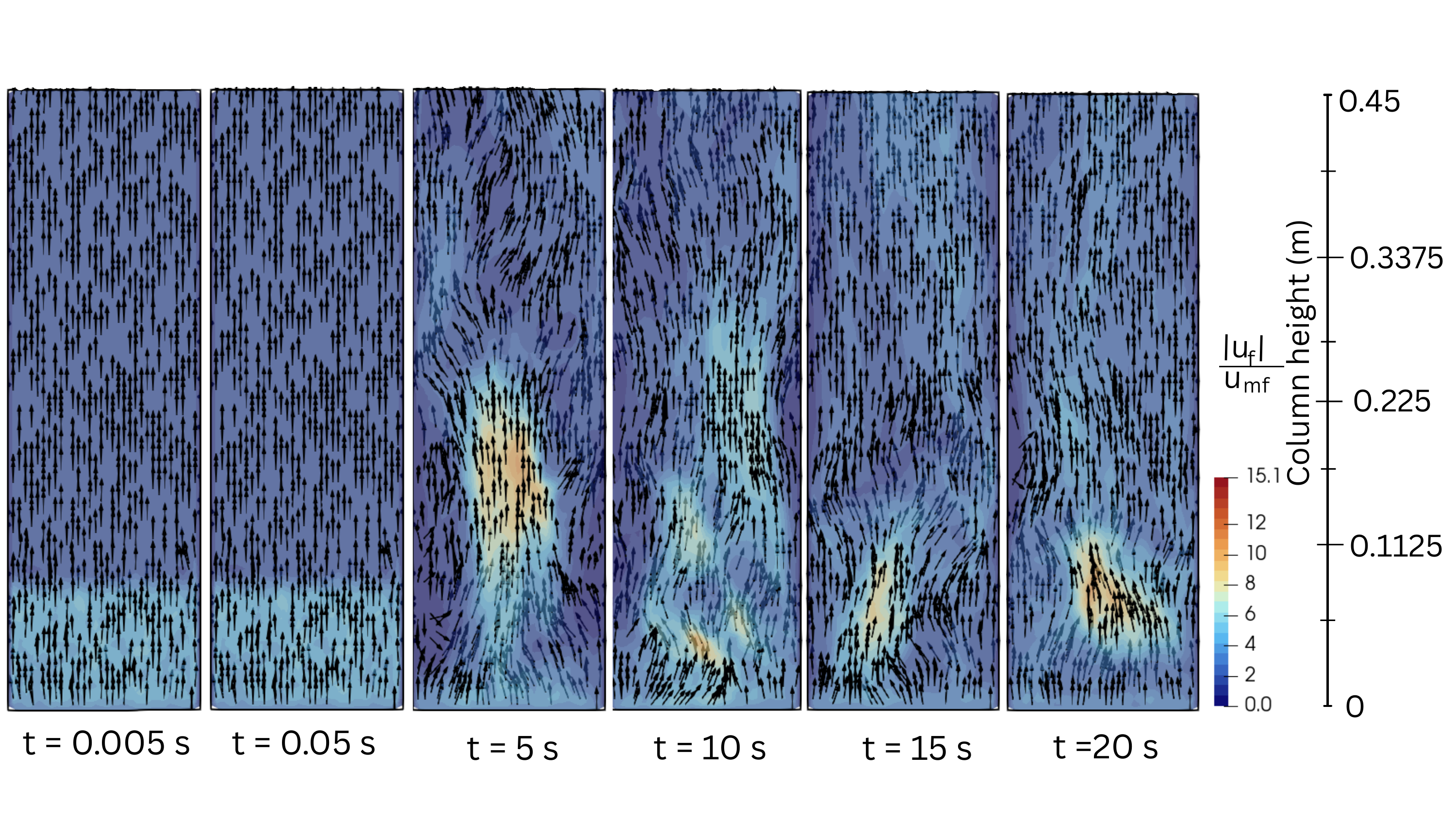}
    \label{fig:air_3umf1}
  }

  \caption{Instantaneous dimensionless fluid velocity fields showing normalized velocity magnitude, $|\mathbf{u}_f|/U_{mf}$ (colormap) and in-plane (projected) velocity vectors on a vertical mid-plane slice ($x = 0$) for (a, c, e) water and (b, d, f) air at $U_s/U_{mf}=2$, $2.5$, and $3$, captured at representative time instants.}
  \label{fig:velfield_snap}

\end{figure*}
The instantaneous velocity fields in Figure~\ref{fig:velfield_snap} provide a spatial view of this transition. Increasing $U_s/U_{mf}$ produces progressively stronger velocity gradients, recirculation, and high-velocity pathways in both media. These structures remain comparatively smooth and coherent in water, whereas air develops sharper gradients, intermittent upward pathways, and greater spatial variability, consistent with the larger fluctuations in Figure~\ref{fig:fluctuation}.

\begin{table}[htbp]
\centering
\caption{Comparison of the principal dimensionless parameters for the water- and air-fluidized beds. The Archimedes and Reynolds numbers were calculated using the Sauter mean diameter of the binary particle mixture and superficial velocity.}
\label{tab:dimensionless}
\begin{tabular}{cccc}
\hline
Fluid & $\rho_p/\rho_f$ & $Ar$ & $Re_s$ range \\
\hline
Water & 2.23 &$4.74 \times 10^{5}$ & 49--294 \\
Air   & 1858.3 & $2.86 \times 10^{6}$ & 142--854 \\
\hline
\end{tabular}
\end{table}

The contrasting dynamics can be interpreted through the governing dimensionless parameters (Table~\ref{tab:dimensionless}). Compared with water, air has a much larger particle-to-fluid density ratio and Archimedes number and operates at higher Reynolds numbers, reflecting a substantially different balance of inertial, viscous, and gravitational effects \cite{glicksman1993simplified, kunii1991fluidization}. Thus, matching $U_s/U_{mf}$ does not imply dynamic similarity between the two media; the weaker damping and greater relative particle inertia in air are consistent with its larger fluctuations and greater flow heterogeneity.

\subsection{Bed expansion}
\label{sec5p3}

Bed expansion was examined through its transient evolution and variation with superficial velocity. The bed height, $H_e$, was determined from the cross-sectionally averaged solid-fraction profile using a threshold of $\alpha_s=0.95$.

\begin{figure*}[t]
\centering
\subfigure[Water]{
\includegraphics[width=0.48\textwidth]{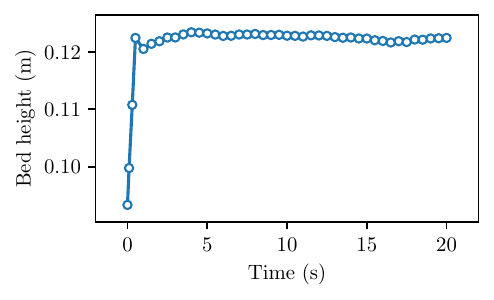}
\label{fig:bed_time_water}
}
\hfill
\subfigure[Air]{
\includegraphics[width=0.48\textwidth]{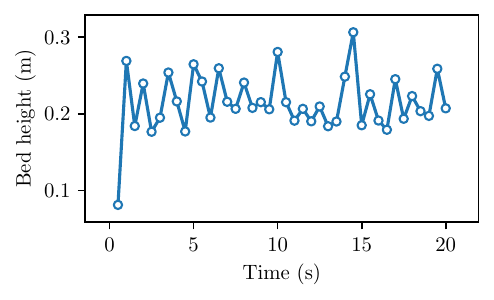}
\label{fig:bed_time_air}
}
\caption{Temporal evolution of bed height at $U_s/U_{mf}=3$ for (a) water and (b) air.}
\label{fig:bed_time}
\end{figure*}

Figure~\ref{fig:bed_time} shows the transient bed height at $U_s/U_{mf}=3$. In water, the bed rapidly approaches a quasi-steady height and subsequently exhibits only small fluctuations, consistent with the well-damped liquid--solid fluidization reported by \citet{islam2023liquid}. In air, substantially larger and persistent oscillations occur due to repeated expansion and collapse associated with gas-void and bubble dynamics. The contrasting responses are consistent with the stronger hydrodynamic damping in water and the more heterogeneous fluidization observed in air.

\begin{figure}[!htbp]
	\centering
	\includegraphics{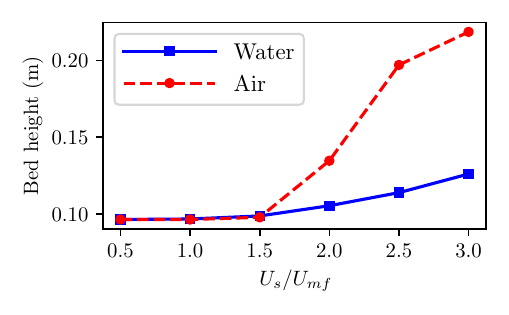}
	\caption{Variation of mean expanded bed height with normalized superficial velocity. A cell size of 5~mm is used here.}
	\label{fig:bed_SV}
\end{figure} 
Figure~\ref{fig:bed_SV} shows that bed expansion increases with $U_s/U_{mf}$ in both media, but with markedly different magnitudes. Expansion in water remains relatively modest and increases gradually, whereas the air-fluidized bed expands sharply at higher velocities. Thus, matching $U_s/U_{mf}$ does not produce comparable bed expansion across the two media, consistent with their different hydrodynamic regimes and fluid properties.

\subsection{Pressure drop}\label{sec5p4}

The bed pressure drop, $\Delta P$, arises from the cumulative loss of fluid momentum due to viscous and inertial interactions with the particles, and therefore provides a direct macroscopic measure of interphase momentum exchange within the bed.
In the present CFD--DEM simulations, $\Delta P$ was calculated as the difference between the cross-sectionally averaged pressure at the inlet and at the bed surface, and subsequently time-averaged in the statistically steady window.
The hydrostatic pressure contribution was excluded from the reported values.

At $U_s=3U_{mf}$, the water-fluidized bed reached a statistically steady pressure drop of $0.651\pm0.007$~kPa over $t=8$--$20$~s, corresponding to a relative fluctuation of only 1.10\%. In contrast, the air-fluidized bed exhibited a mean pressure drop of $1.168\pm0.372$~kPa and a substantially larger relative fluctuation of 31.88\%. The stronger fluctuations in air are consistent with bubble-driven variations in solids concentration and interphase momentum exchange, whereas stronger hydrodynamic damping produces a more stable response in water \cite{peng2022experimental}. This behavior is also consistent with the contrasting bed-height fluctuations discussed in Section~\ref{sec5p3}.

Before fluidization, the pressure drop was compared with the Ergun relation \cite{ergun1952}:
\begin{equation}
\frac{\Delta P}{H_e} =
\frac{150(1-\alpha_f)^2 \mu_f U_s}{\alpha_f^3 d_{32}^2}
+
\frac{1.75(1-\alpha_f)\rho_f U_s^2}{\alpha_f^3 d_{32}},
\label{eq:ergun_pd}
\end{equation}
where the Sauter mean diameter of the binary mixture is $d_{32}=3.4$~mm. The Ergun relation is used here as a theoretical reference for the fixed-bed regime rather than an exact representation of the bidisperse packing \cite{koekemoer2015effect,marchelli2023drag,gao2024modified}. At incipient fluidization, the pressure drop balances the effective weight of the particles,
\begin{equation}
\frac{\Delta P}{H_e}=(\rho_p-\rho_f)(1-\alpha_f)g.
\label{eq:pd_weight}
\end{equation}

\begin{figure*}[!htbp]
	\centering
	\subfigure[Water]{
		\includegraphics{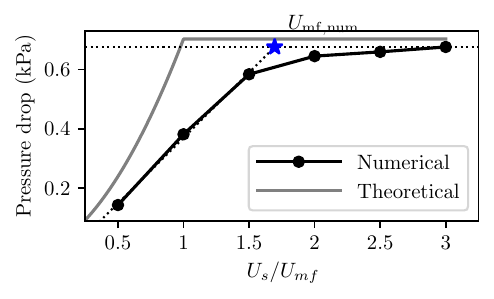}
		\label{fig:pd_us_water}
	}
	\hfill
	\subfigure[Air]{
		\includegraphics{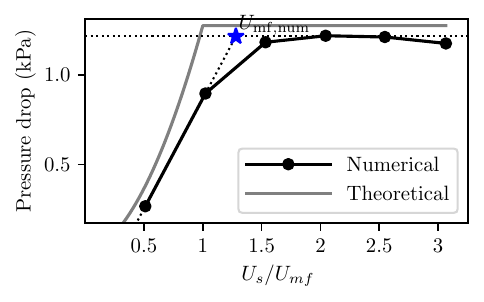}
		\label{fig:pd_us_air}
	}
	\caption{Pressure drop versus superficial velocity from CFD--DEM simulations and the theoretical fixed-bed response for (a) water and (b) air.}
	\label{fig:pd_us}
\end{figure*}
Figure~\ref{fig:pd_us} shows the expected increase in pressure drop in the fixed-bed regime followed by a change in slope as fluidization is established. The numerical minimum fluidization velocity, $U_{mf,\mathrm{num}}$, determined by the tangent-intersection method, is approximately 1.7 times the Wen--Yu estimate for both media. This difference is attributed to the layered bidisperse packing and associated particle interactions, local voidage variations, and wall effects that are not represented in the monodisperse Wen--Yu correlation \cite{alghamdi2021assessment}. Accordingly, $U_{mf,\mathrm{num}}$ provides the effective minimum fluidization velocity of the simulated beds, while the Wen--Yu values are retained only as reference velocities for normalizing the operating conditions.
Beyond minimum fluidization, idealized theory predicts that the bed pressure drop remains approximately constant, since the particle bed is largely supported by the fluid drag. Accordingly, the theoretical reference curves shown in Figure~\ref{fig:pd_us} are constructed using the Ergun relation in the fixed-bed regime and a plateau beyond $U_{mf}$. The  minimum fluidization velocity for the theoretical curve was calculated by equating Eqs.~\eqref{eq:ergun_pd} and \eqref{eq:pd_weight}. 

\begin{table}[h!]
\centering
\caption{Comparison of pressure drop at \emph{minimum fluidization} predicted by theoretical (Eq.~\eqref{eq:pd_weight}) and CFD--DEM (numerical; as shown in Figure~\ref{fig:pd_us}) simulations for water and air.}
\renewcommand{\arraystretch}{1.5} 
\setlength{\tabcolsep}{3pt}      
\begin{tabular}{lccc}
\hline
\textbf{Fluid} & 
\textbf{$\Delta P_{\text{th}}$ [kPa]} & \textbf{$\Delta P_{\text{num}}$ [kPa]} & \textbf{ Relative error [\%]} \\
\hline
\textbf{Water} & 0.703  & 0.675 & 3.9 \\
\textbf{Air}   & 1.273  & 1.213 & 4.7 \\
\hline
\end{tabular}
\label{tab:umf_pressure_velocity}
\end{table}
The theoretical and numerical pressure drops at minimum fluidization agree within 5\% for both media (Table~\ref{tab:umf_pressure_velocity}), showing that the simulations recover the expected macroscopic force balance at incipient fluidization.

\subsection{Particle volume-fraction distribution}
\label{sec5p5}
\begin{figure*}[!htbp]
  \centering

  \subfigure[ Water,  $U_s/U_{mf}=2$]{
    \includegraphics[width=0.47\textwidth]{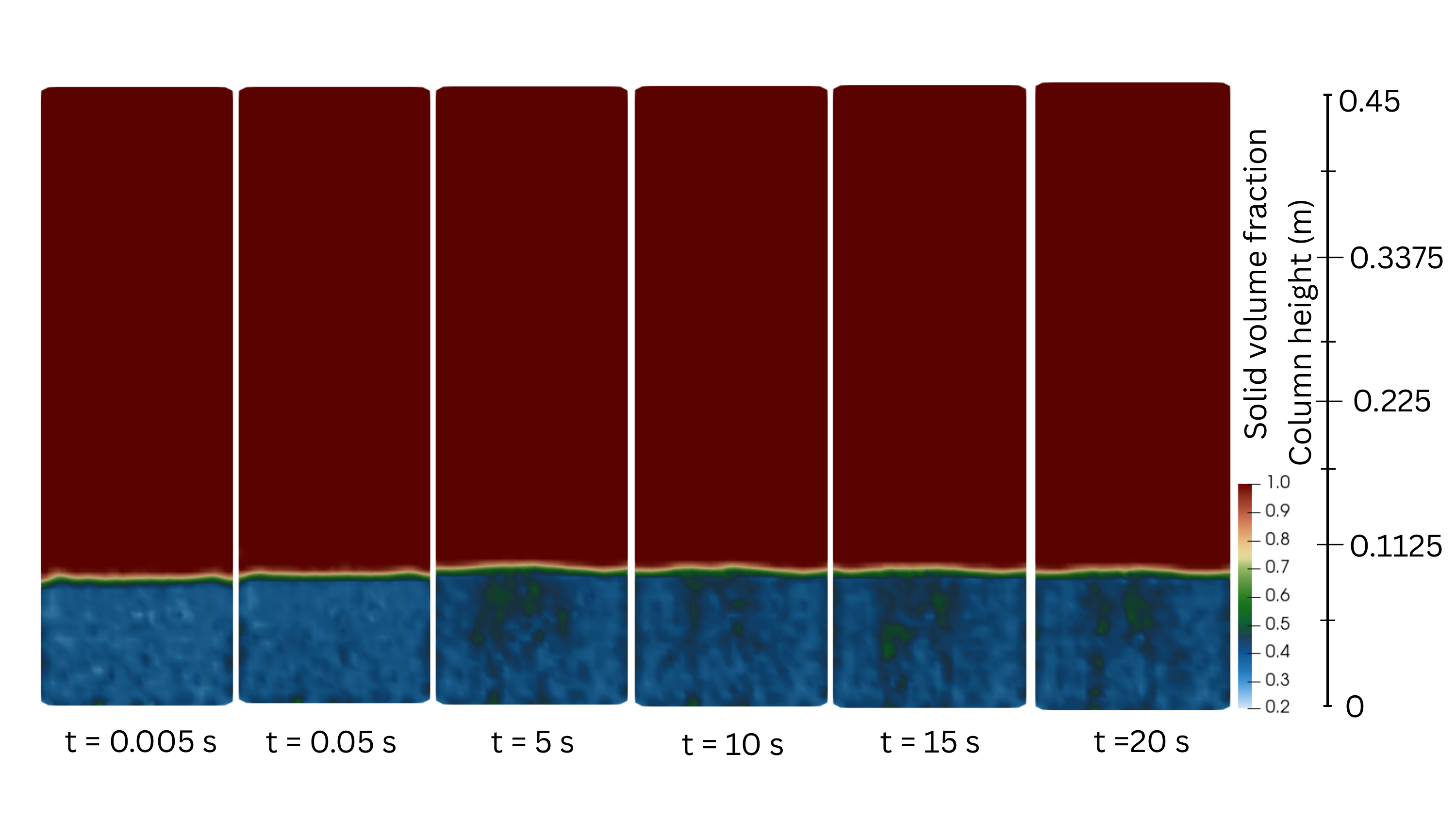}
    \label{fig:water_2umf}
  }
  \hfill
  \subfigure[ Air,  $U_s/U_{mf}=2$]{
    \includegraphics[width=0.47\textwidth]{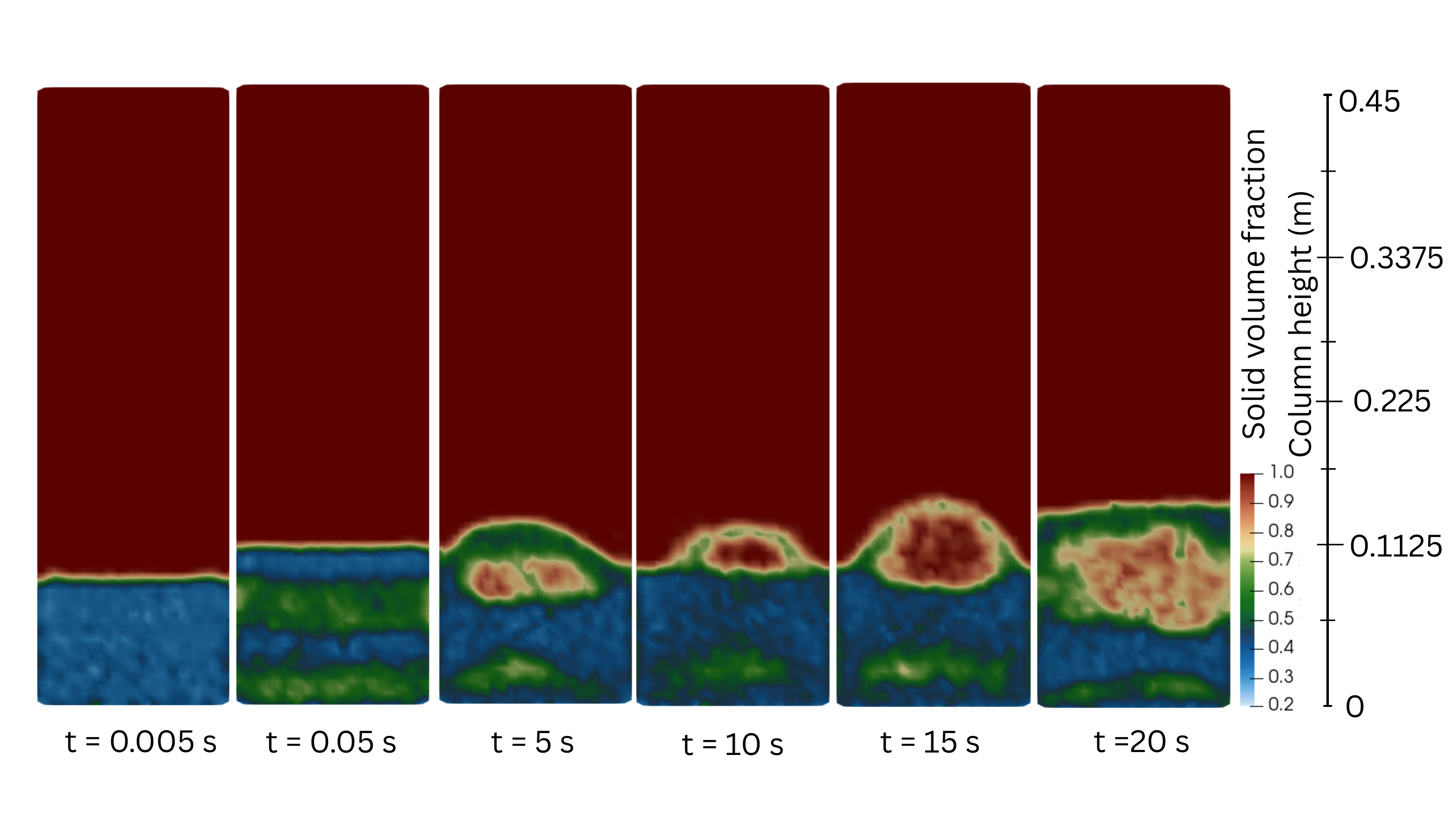}
    \label{fig:air_2umf}
  }

  \subfigure[ Water,  $U_s/U_{mf}=3$]{
    \includegraphics[width=0.47\textwidth]{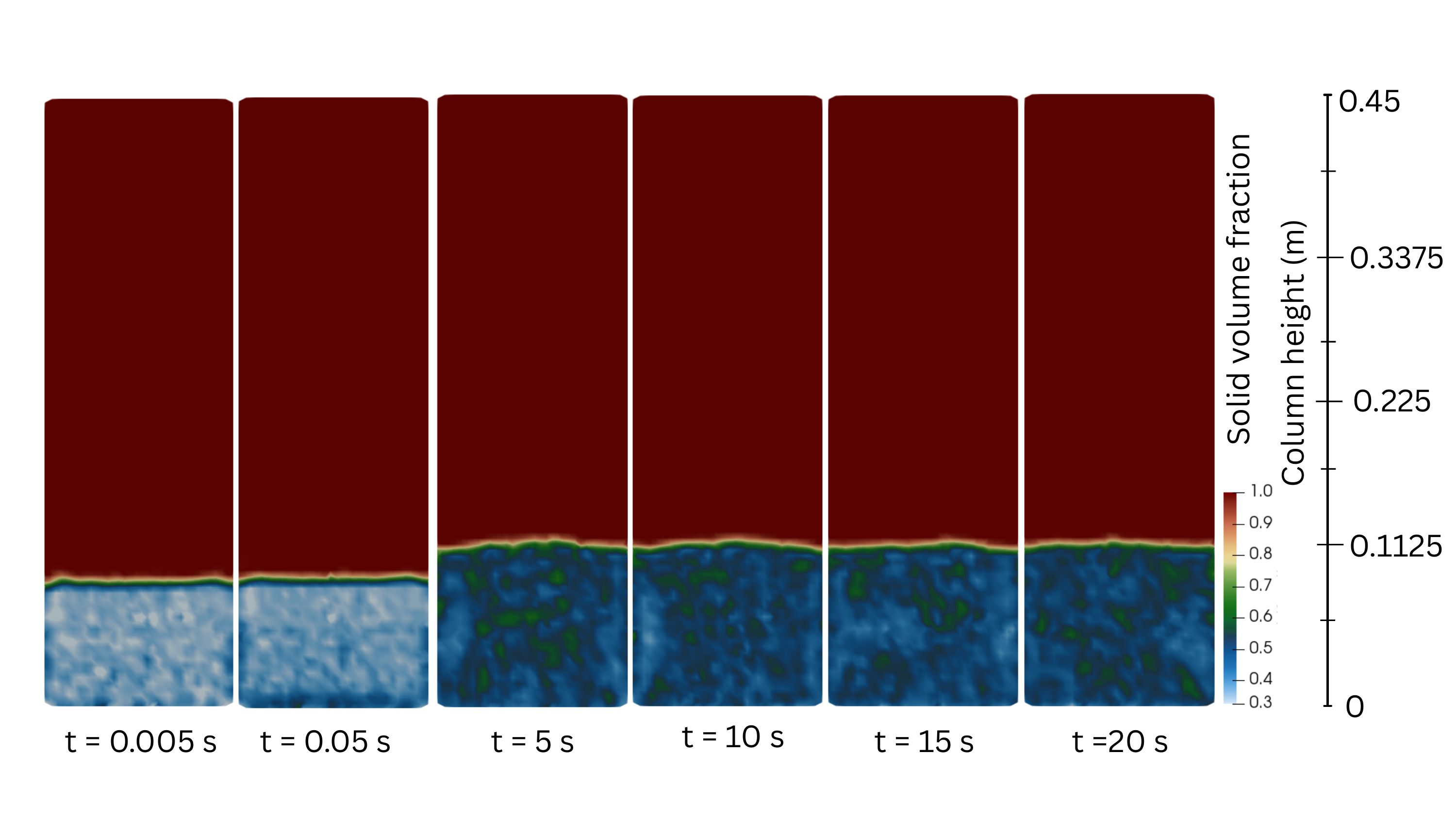}
    \label{fig:water_3umf}
  }
  \hfill
  \subfigure[ Air,  $U_s/U_{mf}=3$]{
    \includegraphics[width=0.47\textwidth]{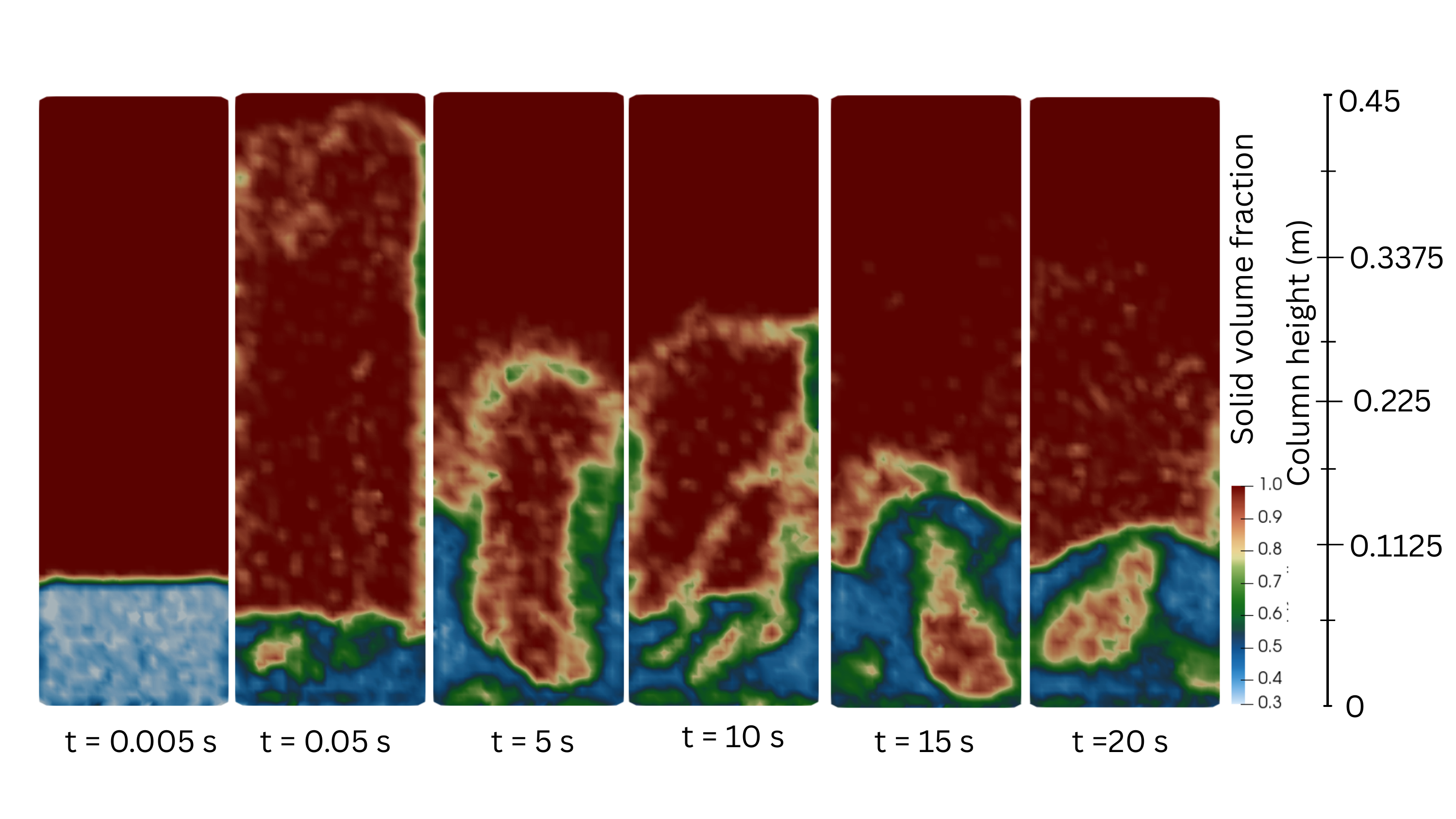}
    \label{fig:air_3umf}
  }  
  \caption{Instantaneous solid volume fraction ($\alpha_s$) fields on vertical mid-plane ($x=0$) for (a, c) water and (b, d) air at $U_s/U_{mf}=2$ and $3$.}
  \label{fig:solidfrac_snap}
\end{figure*}

Figure~\ref{fig:solidfrac_snap} shows the development of solids heterogeneity with increasing superficial velocity. The water-fluidized bed remains comparatively dense and uniform, with a relatively sharp bed interface even at $U_s/U_{mf}=3$. In air, increasing velocity produces progressively larger void-rich regions and a more diffuse bed interface, resulting in substantially greater spatial heterogeneity. These differences are consistent with the stronger hydrodynamic damping in water and the more vigorous gas--solid dynamics identified from the velocity and bed-expansion analyses \cite{peng2022experimental}.

\subsection{Particle dynamics}
\label{sec5p6b}

\begin{figure*}[!htbp]
	\centering
	\subfigure[Water]{
		\includegraphics[width=0.48\textwidth]{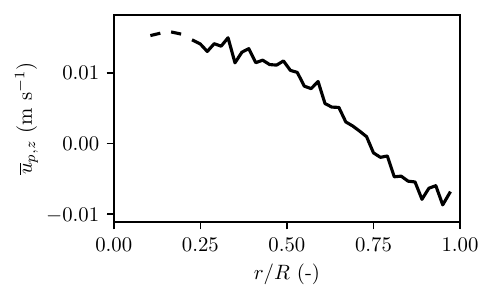}
		\label{fig:upz_rad_water}
	}
	\hfill
	\subfigure[Air]{
		\includegraphics[width=0.48\textwidth]{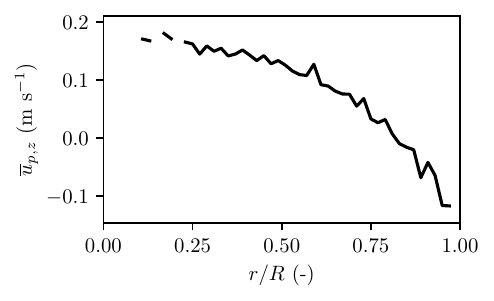}
		\label{fig:upz_rad_air}
	}
	\caption{Time-averaged radial profiles of axial particle velocity at $U_s=3U_{mf}$ for (a) water and (b) air. Positive and negative values
	indicate upward and downward particle motion, respectively. Profiles are obtained from annular radial binning and time averaging over the statistically steady regime (over $t=15$--$20$~s) over the full bed height.}
	\label{fig:upz_rad}
\end{figure*}

Particle transport was characterized using the time-averaged axial particle velocity and its fluctuations at $U_s=3U_{mf}$.
Figure~\ref{fig:upz_rad} shows net upward particle motion in the core and downward motion toward the wall in both media, indicating core--annulus circulation. The circulation is comparatively weak in water but substantially stronger in air, where larger velocity magnitudes indicate more vigorous vertical particle transport. This difference is consistent with the greater bed-scale unsteadiness and heterogeneity observed for the air-fluidized bed.

\begin{figure*}[t!]
	\centering
	\subfigure[Water]{
		\includegraphics[width=0.48\textwidth]{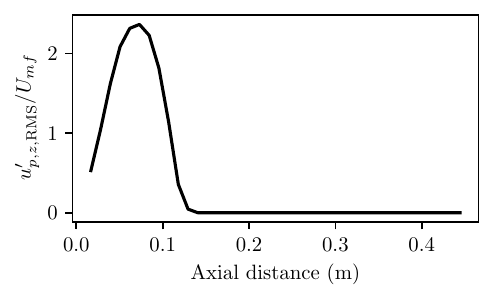}
		\label{fig:upz_rms_water}
	}
	\hfill
	\subfigure[Air]{
		\includegraphics[width=0.48\textwidth]{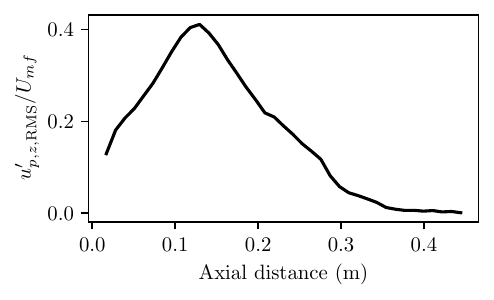}
		\label{fig:upz_rms_air}
	}
	\caption{Axial distribution of the RMS particle axial velocity fluctuation at $U_s=3U_{mf}$ for (a) water and (b) air, evaluated over $t=5$--$20$~s. The RMS fluctuation is calculated as 
    $u'_{p,z,\mathrm{RMS}} = \sqrt{\langle u_{p,z}^{2} \rangle - \langle u_{p,z} \rangle^{2}}$, where averaging is performed over all particles and time within each height bin. 
    Profiles are obtained over the steady-state regime ($t = 5$--$20\,\mathrm{s}$).}
	\label{fig:upz_rms}
\end{figure*}

Figure~\ref{fig:upz_rms} shows substantially greater particle velocity fluctuations in air over most of the bed height. In water, the fluctuations are concentrated within the dense bed and decay rapidly toward the bed surface, whereas in air they remain appreciable over a larger vertical extent. The stronger particle agitation in air is consistent with its bubble-driven, heterogeneous dynamics and with the larger fluid-velocity fluctuations observed earlier. Similar differences between liquid- and gas-fluidized beds have been reported experimentally by \citet{peng2022experimental}.

\subsection{Microstructural analysis of packing and contact forces}
\label{sec5p6c}

\begin{figure*}[!htbp]
	\centering
	\subfigure[]{
		\includegraphics[width=0.48\textwidth]{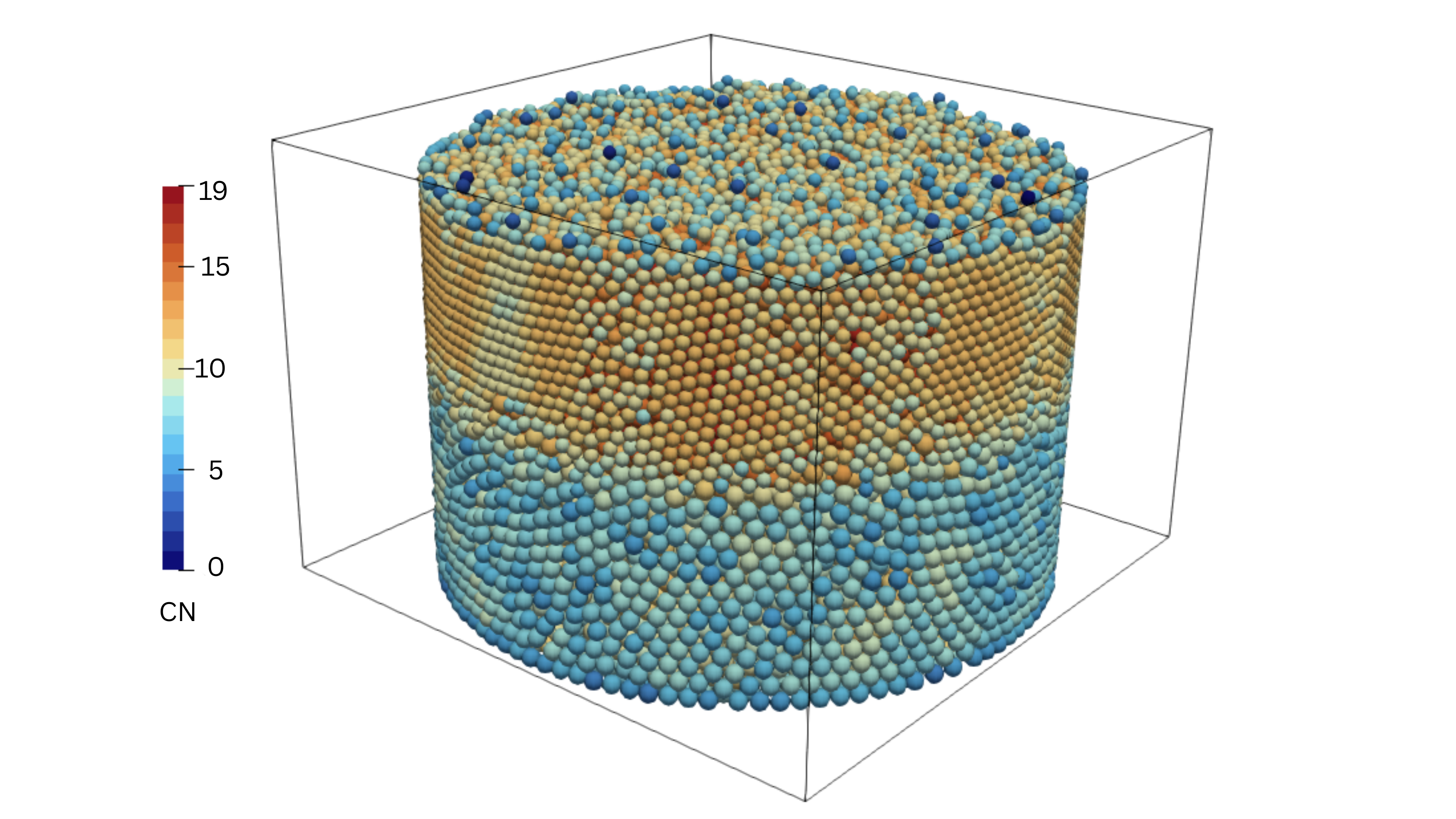}
		\label{fig:cn_snap}
	}
	\hfill
	\subfigure[]{
		\includegraphics[width=0.48\textwidth]{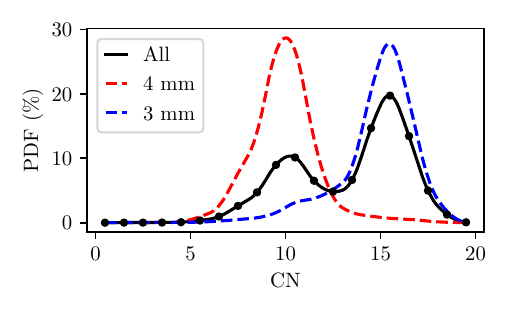}
		\label{fig:pdf_cn}
	}
	\caption{Spatial distribution of coordination number in the initial packing: (a) color-coded representation of CN in the packed bed, and (b) corresponding probability density function of CN, showing a bimodal distribution associated with 3 mm and 4 mm particles.}
	\label{fig:cn}
\end{figure*}

The macroscopic differences in solids distribution and particle transport discussed in the previous subsections originate from changes in the underlying contact network. To quantify these microstructural effects, the coordination number (CN), contact-force statistics, and particle drag-force distributions were examined. Here, the coordination number denotes the number of particle contacts associated with an individual particle.
Figure~\ref{fig:cn} characterizes the initial packed bed before fluidization. 
The spatial distribution reveals that the contact network is not uniform across the bed height, with lower coordination numbers concentrated in the lower region and higher values prevailing in the middle and upper packed zones. This indicates that the observed bimodal PDF is influenced not only by bidispersity, but also by axial variations in local packing structure. 
This behavior differs from conventional observations reported in the literature \cite{zhang2021numerical} for randomly packed systems, where larger particles typically exhibit higher coordination numbers due to their dominant role in the contact network. Overall, the results indicate that the coordination number distribution is significantly influenced by the packing strategy and provides valuable insight into the microstructural characteristics of the packed bed.

\begin{figure*}[!htbp]
	\centering
	\subfigure[Water]{
		\includegraphics[width=0.48\textwidth]{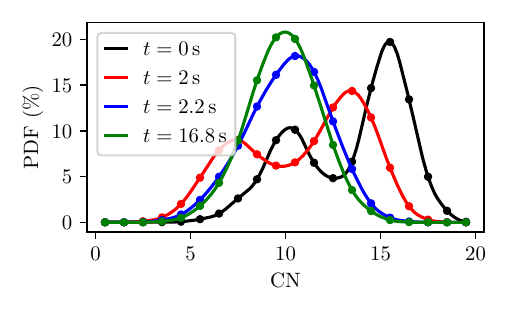}
		\label{fig:cn_water}
	}
	\hfill
	\subfigure[Air]{
		\includegraphics[width=0.48\textwidth]{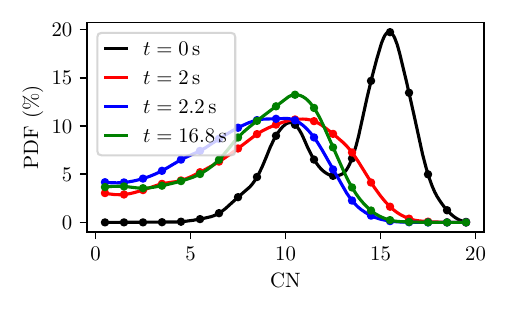}
		\label{fig:cn_air}
	}
	\caption{Temporal evolution of the coordination number distribution, represented by the probability density function, at $t=0$, $2$, $2.2$, and $16.8$ s for $U_s=3U_{mf}$ for: (a) water and (b) air fluidization, illustrating changes in particle contacts during bed expansion.}
	\label{fig:cn2}
\end{figure*}

The temporal evolution of CN during fluidization is shown in Figure~\ref{fig:cn2}. In the water-fluidized bed, the distribution shifts toward lower CN values soon after fluidization begins, indicating progressive breakup of the initially dense contact network as the bed expands and particles become suspended. 
Another notable feature is the progressive disappearance of the initially bimodal CN distribution seen in the packed bed. As fluidization proceeds, the two peaks associated with the different particle sizes merge into a single dominant peak, indicating that particle rearrangement and mixing progressively erase the distinct local contact environments of the 3 mm and 4 mm particles.
At later times, the distribution remains concentrated around moderate CN values, suggesting a comparatively homogeneous suspension.

In contrast, the air-fluidized bed retains a broader CN distribution over time. 
Although the bimodal structure also weakens in the air-fluidized bed, the resulting distribution remains broader and less sharply unimodal than in water, reflecting persistent mesoscale heterogeneity caused by bubbling and transient clustering.
Although the initial dense-contact peak weakens during expansion, appreciable probability remains over a wide CN range, indicating alternating regions of clustering and local voidage associated with bubbling motion \cite{peng2016cfd}. The larger fraction of low-CN or weakly connected particles also reflects transient particle isolation in dilute regions.

\begin{figure*}[!htbp]
	\centering
	\subfigure[Water]{
		\includegraphics[width=0.48\textwidth]{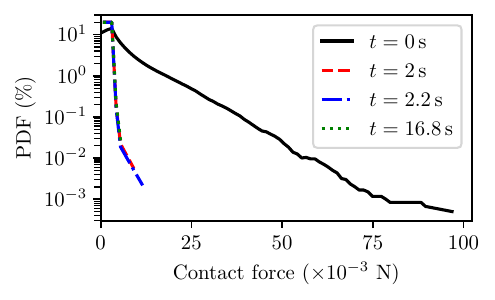}
		\label{fig:cf_water}
	}
	\hfill
	\subfigure[Air]{
		\includegraphics[width=0.48\textwidth]{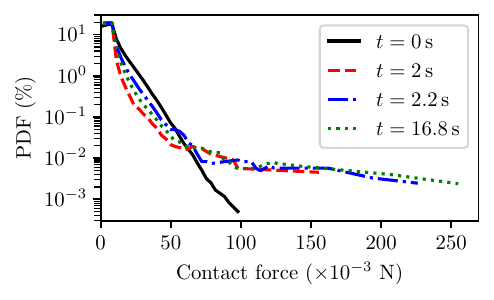}
		\label{fig:cf_air}
	}
	\caption{Temporal evolution of the total contact force, represented by the probability density function, at $t=0$, $2$, $2.2$, and $16.8$ s for $U_s=3U_{mf}$ for: (a) water and (b) air fluidization, illustrating changes in particle contacts during bed expansion.}
    \label{fig:cf}
\end{figure*}

The evolution of the contact network is further reflected in the total contact-force distributions shown in Figure~\ref{fig:cf}. Fluidization substantially reduces contact forces in both media, but the response differs markedly. In water, the distribution rapidly collapses toward low force magnitudes as the initially load-bearing contact network is disrupted. In air, a broader force distribution persists, indicating intermittent formation of stronger contacts during clustering and bed rearrangement. Together with the coordination-number distributions, these results show that water fluidization produces a more uniformly dispersed contact network, whereas air fluidization sustains intermittent particle clustering and contact formation.

\begin{figure*}[t!]
	\centering
	\subfigure[Water]{
		\includegraphics[width=0.48\textwidth]{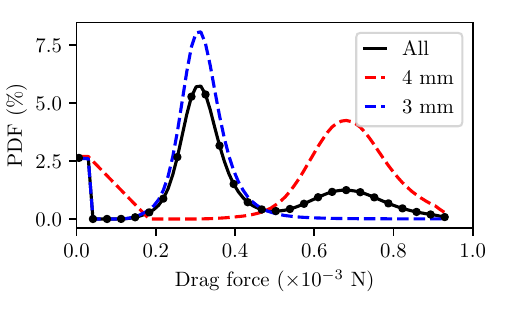}
		\label{fig:drag_wa}
	}
	\hfill
	\subfigure[Air]{
		\includegraphics[width=0.48\textwidth]{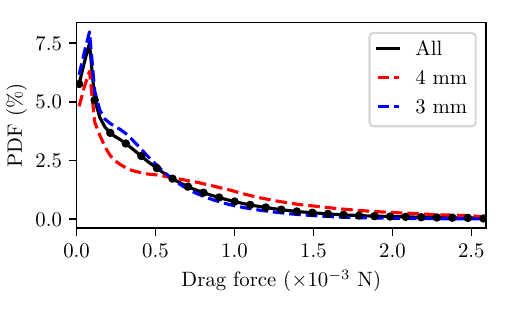}
		\label{fig:drag_ai}
	}
	\caption{Probability density function of drag force acting on particles, obtained by aggregating data over multiple timesteps at $U_s=3U_{mf}$ for (a) water and (b) air fluidization.}
	\label{fig:drag}
\end{figure*}

Finally, Figure~\ref{fig:drag} compares the hydrodynamic drag experienced by particles in the two media. Because drag depends on fluid density, local void fraction, and slip velocity (Eq.~\eqref{eq:drag_def}), it provides a direct measure of fluid--particle coupling strength.
The water-fluidized bed exhibits a bimodal drag-force distribution, indicating stronger particle-size-dependent fluid--particle interaction. This is consistent with the higher density and viscosity of water, which enhance momentum transfer to the solids phase. In contrast, the air-fluidized bed shows a pronounced peak at low drag forces followed by a smoother decay, reflecting weaker average drag but larger mesoscale heterogeneity arising from bubbles and void regions. Moreover, the dependence of particle size is weaker than in water. These microstructural observations help explain why water fluidization produces a more homogeneous and well-damped bed, whereas air fluidization leads to stronger fluctuations, intermittent contacts, and heterogeneous transport.

The coordination-number and contact-force distributions show that the two fluidizing media produce distinctly different particle-scale interaction networks. In water, fluidization rapidly disrupts the initial contact network and leaves predominantly weak particle contacts, whereas air sustains intermittent clustering and stronger contact interactions. These differences in particle-scale dynamics ultimately govern the rearrangement and transport of the two particle species. The resulting mixing behavior is examined quantitatively in the following subsection.

\subsection{Particle mixing}\label{sec5p6a}

Mixing of the binary particle bed was quantified using the Lacey mixing index (LMI) \cite{lacey1954developments}, which provides a normalized measure of the degree of mixing. The index varies from 0 (fully segregated) to 1 (perfectly mixing limit) defined by Lacey formulation. In the present simulations, the evolution of mixing process is also qualitatively illustrated by the progressive interpenetration of the initially layered particle configuration in Figures~\ref{fig:7} and~\ref{fig:8}.

To compute the LMI, the domain was partitioned into cubic sampling cells, with only cells containing at least three particles considered valid.
Within each valid cell, the local composition of the reference species, chosen here as the large particles, was defined as:
\begin{equation}
\phi_i = \frac{n_{L,i}}{n_{i}},
\end{equation}
where $n_{L,i}$ and $n_i$ are the numbers of large and total particles, respectively, in cell $i$. 
The spatial variance of composition is:
\begin{equation}
S^2 = \frac{1}{N-1}\sum_{i=1}^{N}\left(\phi_i-\phi_m\right)^2,
\end{equation}
where $N$ is the number of valid sampling cells and $\phi_m$ is the domain-averaged composition of the reference species. The Lacey mixing index is then defined as:
\begin{equation}
\mathrm{LMI}=\frac{S_0^2-S^2}{S_0^2-S_m^2},
\label{eq:lmi}
\end{equation}
where $S_0^2=\phi_m(1-\phi_m)$ and $S_m^2=\phi_m(1-\phi_m)/\bar n$ represent the variances for the completely segregated and randomly mixed states, respectively, with $\bar n$ denoting the mean occupancy of the valid cells.
Although particle occupancy varies among individual sampling cells, the random-mixing variance in the classical Lacey formulation is evaluated using the mean particle occupancy, $\bar{n}$, of the valid sampling cells. To reduce statistical fluctuations associated with sparsely populated cells, only sampling cells containing at least three particles $(n \geq 3)$ were included in the calculation.

\begin{figure*}[t]
    \centering
    \subfigure[Water]{
        \includegraphics[width=0.48\textwidth]{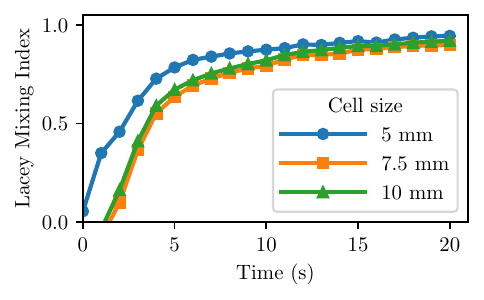}
        \label{fig:lmi_time_water}
    }
    \hfill
    \subfigure[Air]{
        \includegraphics[width=0.48\textwidth]{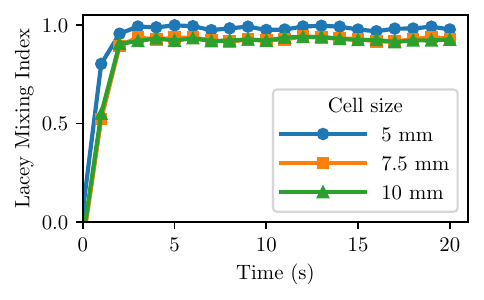}
        \label{fig:lmi_time_air}
    }
    \caption{Temporal evolution of the Lacey mixing index at $U_s=3U_{mf}$ for (a) water and (b) air fluidization.}
    \label{fig:lmi_time}
\end{figure*}

\begin{table}[ht]
\centering
\caption{Cell-size sensitivity analysis of the classical Lacey Mixing Index for the water- and air-fluidized beds (steady-state averages.}
\begin{tabular}{
    >{\centering\arraybackslash}p{1.0cm}
    c
    c
    >{\centering\arraybackslash}p{1.0cm}
    >{\centering\arraybackslash}p{1.0cm}
    >{\centering\arraybackslash}p{1.4cm}
}
\hline
\textbf{Medium} &
\textbf{Cell size} &
\textbf{Mean} &
\textbf{Valid cells,} &
\textbf{Mean occupancy,}  &
\textbf{Rejected occupied} \\
 &
\textbf{(mm)} &
\textbf{LMI}
 &
 $N$
 &
 $\bar{n}$
 &
\textbf{cells (\%)} \\
\hline
Water & 5.0  & 0.910 & 9739 & 3.74  & 34.16 \\
       & 7.5 & 0.855 & 4553 & 9.85  & 3.01  \\
       &10.0 & 0.878 & 1986 & 22.64 & 3.30  \\
\hline
Air    & 5.0 & 0.964 & 7507 & 3.62  & 61.95 \\
       & 7.5 & 0.931 & 5521 & 7.54  & 28.81 \\
       &10.0 & 0.927 & 2999 & 14.59 & 20.25 \\
\hline
\end{tabular}
\label{tab:cellsize}
\end{table}
A cell-size sensitivity analysis was performed using cell sizes of 5, 7.5 and 10 mm and is summarized in Figure~\ref{fig:lmi_time} and Table~\ref{tab:cellsize}. For both media, the 5 mm sampling cells exhibited the lowest mean particle occupancy and the highest fraction of rejected occupied cells. Increasing the cell size to 7.5 and 10 mm substantially increased the mean particle occupancy and reduced the proportion of rejected cells. Despite these differences in cell occupancy, the steady-state LMI values obtained using the 7.5 and 10 mm sampling cells differed only marginally for both media, indicating that the classical LMI becomes essentially independent of the sampling-cell size once sufficient particle occupancy is achieved.

Figure~\ref{fig:lmi_time} compares the transient evolution of LMI at $U_s=3U_{mf}$ for the two media. In the water-fluidized bed (Figure~\ref{fig:lmi_time_water}), LMI increases gradually and approaches an asymptotic value over several seconds, consistent with a comparatively damped, dense-phase motion. In contrast, the air-fluidized bed (Figure~\ref{fig:lmi_time_air}) exhibits a rapid initial rise in LMI, reaching a near-mixed state much earlier. The faster initial mixing in air is consistent with the stronger bubble-driven particle agitation and larger particle transport observed in the preceding analyses.

\begin{figure}[!htbp]
    \centering
    \includegraphics{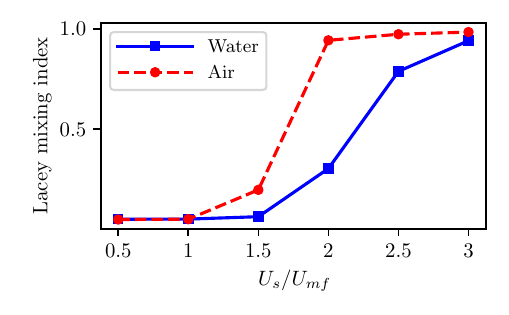}
    \caption{Variation of the time-averaged Lacey mixing index with normalized superficial velocity for water and air. Averages are computed over $t=15$--$20$~s.}
    \label{fig:lmi_vs_us}
\end{figure}
The dependence of mixing on operating condition is summarized in Figure~\ref{fig:lmi_vs_us}, where LMI is time-averaged over $t=15$--$20$~s to represent the statistically steady regime. For both media, mixing improves with increasing superficial velocity, reflecting enhanced particle mobility and stronger convective transport. However, the onset of significant mixing occurs at lower $U_s/U_{mf}$ in air than in water, indicating that gas--solid fluidization achieves strong particle dispersion at comparatively lower normalized velocities. Even at matched $U_s/U_{mf}$, the two media do not collapse to identical mixing levels, highlighting that fluid properties influence not only the fluidization regime but also the underlying mixing mechanism in the binary bed.
The LMI trends quantify the global behavior of particle transport in the bed. To connect these Eulerian statistics to the underlying Lagrangian (local) motion, the representative trajectories of neighboring particles of both sizes in each medium are examined next.

\begin{figure*}[t!]
    \centering
    \subfigure[Water-fluidized bed]{
        \includegraphics[width=0.48\textwidth]{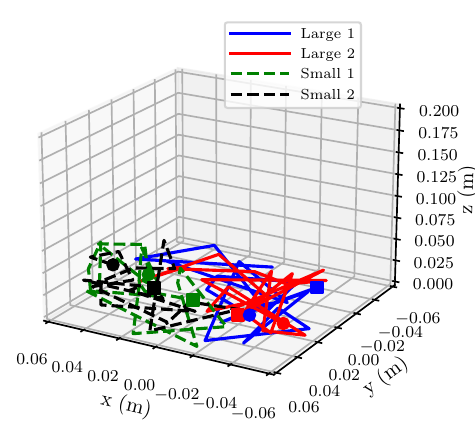}
        \label{fig:traj_water}
    }
    \hfill
    \subfigure[Air-fluidized bed]{
        \includegraphics[width=0.48\textwidth]{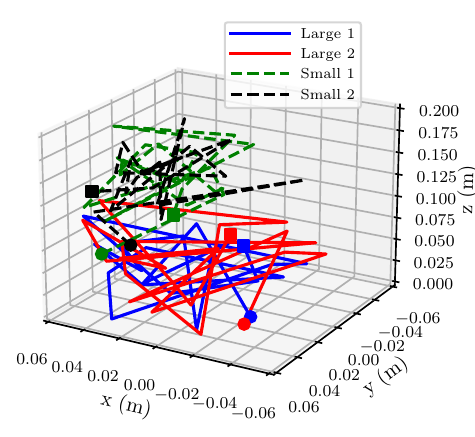}
        \label{fig:traj_air}
    }
    \caption{Three-dimensional trajectories of selected particles at $U_s = 3U_{mf}$ in (a) a water-fluidized bed and (b) an air-fluidized bed. Two small and two large particles that are initially neighboring were tracked using identical particle IDs in both cases, enabling a direct comparison of Lagrangian motion at matched operating conditions. Circles denote initial particle positions and squares denote final positions.}
    \label{fig:traj_combined}
\end{figure*}

For the trajectory analysis, the same two neighboring small particles and two neighboring large particles were tracked in both media, enabling a direct comparison of Lagrangian motion at matched $U_s/U_{mf}$. Figure~\ref{fig:traj_combined} shows that particle paths in the water-fluidized bed remain comparatively confined, with smaller vertical transport and weaker lateral dispersion. 
This behavior reflects the stronger hydrodynamic damping in the liquid phase, which restricts particle displacement and results in weaker mixing and slower homogenization of the bed.
In contrast, the air-fluidized bed exhibits markedly longer trajectories, including larger vertical reach and broader lateral spreading, particularly for the smaller particles. These extended trajectories reflect bubble/void-driven transport and weaker viscous damping, which together enhance particle rearrangement and accelerate the approach to a mixed state. 
The trajectories also suggest greater size-dependent particle mobility in air, particularly for the smaller particles.

\begin{figure*}[!htbp]
    \centering
    \subfigure[Water]{
        \includegraphics[width=0.48\textwidth]{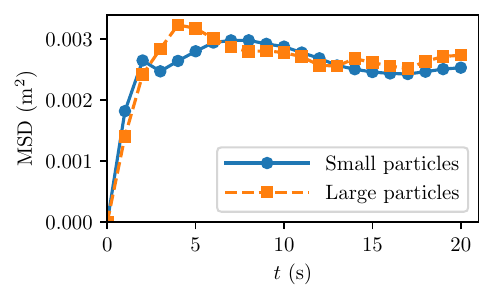}
        \label{fig:msd_water}
    }
    \hfill
    \subfigure[Air]{
        \includegraphics[width=0.48\textwidth]{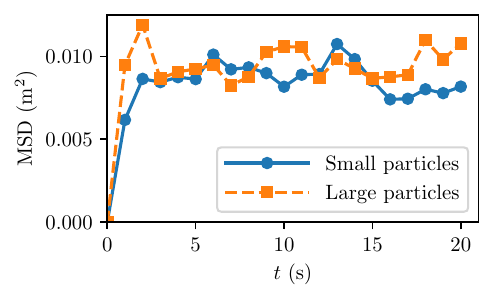}
        \label{fig:msd_air}
    }
    \caption{Time evolution of the mean square displacement (MSD) of the small and large particles in the (a) water-fluidized bed and (b) air-fluidized bed at a superficial velocity of $U_s = 3U_{mf}$. The MSD is evaluated with respect to the particle positions at t=0.}
    \label{fig:msd}
\end{figure*}

To further quantify particle transport, the mean square displacement (MSD) was calculated separately for the small and large particles using all particles of each size class as
\begin{equation}
\mathrm{MSD}_{k}(t)=
\frac{1}{N_{k}}
\sum_{i=1}^{N_{k}}
\left[
\left(x_{i}(t)-x_{i}(0)\right)^{2}
+
\left(y_{i}(t)-y_{i}(0)\right)^{2}
+
\left(z_{i}(t)-z_{i}(0)\right)^{2}
\right],
\label{eq:MSD}
\end{equation}
where $N_{k}$ is the number of particles in particle class $k$, and $(x_i(t),\,y_i(t),\,z_i(t))$ and $(x_i(0),\,y_i(0),\,z_i(0))$
denote the coordinates of particle $i$ at time $t$ and at the initial time, respectively. Figure~\ref{fig:msd} shows that the MSD increases rapidly during the initial stage and subsequently reaches a quasi-steady value in both media. 
The air-fluidized bed exhibits consistently higher MSD values than the water-fluidized bed, indicating enhanced particle transport and dispersion under gas fluidization. In contrast, the lower MSD observed in the water-fluidized bed reflects more restricted particle motion due to stronger liquid-phase damping. These observations are consistent with the trajectory analysis and the faster increase in the Lacey mixing index shown in Figure~\ref{fig:lmi_time}.

Taken together, the LMI, particle trajectories, and MSD results demonstrate that the contrasting hydrodynamic regimes directly influence the mixing of the binary bed. The stronger fluctuations and bubble-driven transport in air produce greater particle mobility and more rapid rearrangement of the initially layered bed, whereas stronger hydrodynamic damping in water restricts particle displacement and slows mixing. Thus, even at matched
$U_s/U_{mf}$, the fluidizing medium governs not only the macroscopic bed response but also the rate and extent of particle mixing.

\section{Conclusions}\label{sec6}

This study used CFD--DEM to compare the hydrodynamics and particle mixing of a bidisperse fluidized bed using water and air under identical geometry, particle properties, and initial conditions. The comparison was performed at matched normalized superficial velocities, $U_s/U_{mf}$, to examine how the fluidizing medium influences both macroscopic bed response and particle-scale dynamics.

At matched $U_s/U_{mf}$, water fluidization produced a comparatively dense and homogeneous bed with stable bed height and pressure drop and weak velocity fluctuations. In contrast, air fluidization generated greater bed expansion, persistent fluctuations, void- and bubble-like structures, and stronger spatial heterogeneity. Particle-velocity and microstructural analyses further showed stronger circulation and a more heterogeneous contact network in air, whereas stronger hydrodynamic damping in water restricted particle motion and promoted a more uniform suspension.

These differences directly influenced particle mixing. The Lacey mixing index increased with superficial velocity in both media, but mixing developed more rapidly and reached higher levels in air at matched $U_s/U_{mf}$. Particle trajectories and mean-square displacement further showed greater particle mobility and transport in air, consistent with its bubble-driven and strongly fluctuating dynamics. Thus, normalization by $U_s/U_{mf}$ does not collapse the hydrodynamic or mixing behavior across the two media; fluid properties remain important in determining the particle-transport mechanisms governing mixing in binary fluidized beds.

The present study is limited to a single column geometry, one particle-size pair with equal particle density, and the Di Felice drag closure. Future work should examine broader size and density ratios, alternative interphase closures, and larger-scale configurations to establish the generality of the observed medium-dependent mixing behavior.

\nomenclature[A001]{$Ar$}{Archimedes number \nomunit{--}}
\nomenclature[A002]{$C_d$}{Drag coefficient \nomunit{--}}
\nomenclature[A003]{$C_o$}{Courant number \nomunit{--}}
\nomenclature[A004]{$D$}{Column diameter \nomunit{\si{\meter}}}
\nomenclature[A005]{$d_i$}{Particle diameter \nomunit{\si{\meter}}}
\nomenclature[A0055]{$d_{32}$}{Sauter mean diameter \nomunit{\si{\meter}}}
\nomenclature[A006]{$d_1$}{Small particle diameter \nomunit{\si{\meter}}}
\nomenclature[A007]{$d_2$}{Large particle diameter \nomunit{\si{\meter}}}
\nomenclature[A008]{$E$}{Young's modulus \nomunit{\si{\pascal}}}
\nomenclature[A009]{$e$}{Restitution coefficient \nomunit{--}}
\nomenclature[A010]{$\mathbf{F}_{c}$}{Contact force \nomunit{\si{\newton}}}
\nomenclature[A011]{$\mathbf{F}_{d}$}{Fluid drag force \nomunit{\si{\newton}}}
\nomenclature[A012]{$G$}{Shear modulus \nomunit{\si{\pascal}}}
\nomenclature[A013]{$\mathbf{g}$}{Acceleration due to gravity \nomunit{\si{\meter\per\second\squared}}}
\nomenclature[A014]{$H$}{Column height \nomunit{\si{\meter}}}
\nomenclature[A015]{$H_b$}{Steady height of the expanded bed \nomunit{\si{\meter}}}
\nomenclature[A015]{$H_e$}{Height of the expanded bed \nomunit{\si{\meter}}}
\nomenclature[A016]{$H_i$}{Initial bed height \nomunit{\si{\meter}}}
\nomenclature[A017]{$\mathbf{I}_i$}{Moment of inertia of particle $i$ \nomunit{\si{\kilogram\metre\squared}}}
\nomenclature[A018]{$\mathbf{K}_{sl}$}{Momentum exchange term 
\nomunit{\si{\newton\per\cubic\metre}}}
\nomenclature[A019]{$m_i$}{Mass of particle $i$ \nomunit{\si{\kilogram}}}
\nomenclature[A020]{$N$}{Total number of cells \nomunit{--}}
\nomenclature[A021]{$p$}{Hydrodynamic (hydrostatic-reduced) fluid pressure \nomunit{\si{\pascal}}}
\nomenclature[A022]{$Re_p$}{Particle Reynolds number \nomunit{--}}
\nomenclature[A023]{$r_{min}$}{Radius of the smallest particle \nomunit{\si{\meter}}}
\nomenclature[A024]{$S^2$}{Variance of particle concentration \nomunit{--}}
\nomenclature[A025]{$S_0^2$}{Variance in completely segregated state \nomunit{--}}
\nomenclature[A026]{$S_m^2$}{Variance in completely mixed state \nomunit{--}}
\nomenclature[A027]{$t$}{Time \nomunit{\si{\second}}}
\nomenclature[A028]{$T_{c}$}{Net moment due to tangential forces \nomunit{\si{\newton\metre}}}
\nomenclature[A029]{$T_{fi}$}{Moment induced by fluid velocity gradient \nomunit{\si{\newton\metre}}}
\nomenclature[A030]{$\mathbf{u}_f$}{Fluid velocity \nomunit{\si{\meter\per\second}}}
\nomenclature[A031]{$\mathbf{u}_i$}{Translational velocity of particle $i$ \nomunit{\si{\meter\per\second}}}
\nomenclature[A032]{$\mathbf{U}_r$}{Relative velocity between solid and fluid \nomunit{\si{\meter\per\second}}}
\nomenclature[A033]{$U_{mf}$}{Minimum fluidization velocity \nomunit{\si{\meter\per\second}}}
\nomenclature[A034]{$U_s$}{Superficial velocity \nomunit{\si{\meter\per\second}}}
\nomenclature[A035]{$V_c$}{Volume of the calculation cell \nomunit{\si{\meter\cubed}}}
\nomenclature[A036]{$U_{mf,\mathrm{num}}$}{Simulation minimum fluidization velocity \nomunit{\si{\meter\per\second}}}

\nomenclature[G101]{$\alpha_f$}{Volume fraction of fluid \nomunit{--}}
\nomenclature[G102]{$\alpha_s$}{Initial solid holdup \nomunit{--}}
\nomenclature[G103]{$\beta$}{Interphase momentum exchange coefficient \nomunit{--}}
\nomenclature[G104]{$\Delta P$}{Pressure drop \nomunit{\si{\pascal}}}
\nomenclature[G105]{$\Delta t_{CFD}$}{CFD time step \nomunit{\si{\second}}}
\nomenclature[G106]{$\Delta t_{DEM}$}{DEM time step \nomunit{\si{\second}}}
\nomenclature[G107]{$\Delta t_{max}$}{Maximum permissible DEM time step \nomunit{\si{\second}}}
\nomenclature[G108]{$\Delta x$}{Mesh size \nomunit{\si{\meter}}}
\nomenclature[G109]{$\mu_a$}{Air viscosity \nomunit{\si{\pascal\second}}}
\nomenclature[G110]{$\mu_f$}{Fluid viscosity \nomunit{\si{\pascal\second}}}
\nomenclature[G111]{$\mu_r$}{Rolling friction coefficient \nomunit{--}}
\nomenclature[G112]{$\mu_s$}{Sliding friction coefficient \nomunit{--}}
\nomenclature[G113]{$\mu_w$}{Water viscosity \nomunit{\si{\pascal\second}}}
\nomenclature[G114]{$\nu$}{Poisson's ratio \nomunit{--}}
\nomenclature[G115]{$\rho_a$}{Air density \nomunit{\si{\kilogram\per\cubic\metre}}}
\nomenclature[G116]{$\rho_f$}{Fluid density \nomunit{\si{\kilogram\per\cubic\metre}}}
\nomenclature[G117]{$\rho_p$}{Particles density \nomunit{\si{\kilogram\per\cubic\metre}}}
\nomenclature[G118]{$\rho_w$}{Water density \nomunit{\si{\kilogram\per\cubic\metre}}}
\nomenclature[G119]{$\tau_f$}{Fluid shear stress tensor \nomunit{\si{\pascal}}}
\nomenclature[G120]{$\phi_i$}{Local concentration of reference particles \nomunit{--}}
\nomenclature[G121]{$\phi_m$}{Mean concentration of reference particles \nomunit{--}}
\nomenclature[G122]{$\boldsymbol{\omega}_i$}{Angular velocity of particle $i$ \nomunit{\si{\radian\per\second}}}

\nomenclature[S201]{$i$}{Particle index}
\nomenclature[S202]{$f$}{Fluid phase}
\nomenclature[S203]{$s$}{Solid phase}
\nomenclature[S204]{$c$}{Contact}
\nomenclature[S205]{$d$}{Drag}
\nomenclature[S206]{$e$}{Expanded state}
\nomenclature[S207]{$mf$}{Minimum fluidization condition}
\nomenclature[S208]{$num$}{Numerical (simulation-based)}
\nomenclature[S209]{$th$}{Theoretical}

\printnomenclature

\vspace{0.5cm}
\noindent
\textbf{Achnowledgements}\\
\noindent RN acknowledges the Ministry of Education (MoE), Government of India, for PhD scholarship support. The authors also acknowledge NSM for computing resources on `PARAM Himalaya' at IIT Mandi, implemented by C-DAC and supported by MeitY and DST, Government of India.\\

\section*{Data availability}

The data that support the findings of this study are available from the
corresponding author upon reasonable request.

\bibliographystyle{elsarticle-harv}
\bibliography{references}

\begin{thebibliography}{36}
\expandafter\ifx\csname natexlab\endcsname\relax\def\natexlab#1{#1}\fi
\providecommand{\url}[1]{\texttt{#1}}
\providecommand{\href}[2]{#2}
\providecommand{\path}[1]{#1}
\providecommand{\DOIprefix}{doi:}
\providecommand{\ArXivprefix}{arXiv:}
\providecommand{\URLprefix}{URL: }
\providecommand{\Pubmedprefix}{pmid:}
\providecommand{\doi}[1]{\href{http://dx.doi.org/#1}{\path{#1}}}
\providecommand{\Pubmed}[1]{\href{pmid:#1}{\path{#1}}}
\providecommand{\bibinfo}[2]{#2}
\ifx\xfnm\relax \def\xfnm[#1]{\unskip,\space#1}\fi
\bibitem[{Alghamdi et~al.(2021)Alghamdi, Peng, Almutairi, Alibrahim, Al-Alweet,
  Moghtaderi and Doroodchi}]{alghamdi2021assessment}
\bibinfo{author}{Alghamdi, Y.A.}, \bibinfo{author}{Peng, Z.},
  \bibinfo{author}{Almutairi, Z.}, \bibinfo{author}{Alibrahim, H.},
  \bibinfo{author}{Al-Alweet, F.M.}, \bibinfo{author}{Moghtaderi, B.},
  \bibinfo{author}{Doroodchi, E.}, \bibinfo{year}{2021}.
\newblock \bibinfo{title}{Assessment of correlations for minimum fluidization
  velocity of binary mixtures of particles in gas fluidized beds}.
\newblock \bibinfo{journal}{Powder Technol.} \bibinfo{volume}{398},
  \bibinfo{pages}{117052. \url{https://doi.org/10.1016/j.powtec.2021.09.035}}.
\bibitem[{Bai et~al.(2023)Bai, Zhao, Lv and Zhou}]{bai2023gas}
\bibinfo{author}{Bai, L.}, \bibinfo{author}{Zhao, Z.}, \bibinfo{author}{Lv,
  W.}, \bibinfo{author}{Zhou, L.}, \bibinfo{year}{2023}.
\newblock \bibinfo{title}{Gas--solid flow characteristics of fluidized bed with
  binary particles}.
\newblock \bibinfo{journal}{Powder Technol.} \bibinfo{volume}{416},
  \bibinfo{pages}{118206. \url{https://doi.org/10.1016/j.powtec.2022.118206}}.
\bibitem[{Clarke et~al.(2018)Clarke, Sederman, Gladden and
  Holland}]{clarke2018investigation}
\bibinfo{author}{Clarke, D.A.}, \bibinfo{author}{Sederman, A.J.},
  \bibinfo{author}{Gladden, L.F.}, \bibinfo{author}{Holland, D.J.},
  \bibinfo{year}{2018}.
\newblock \bibinfo{title}{Investigation of void fraction schemes for use with
  {CFD--DEM} simulations of fluidized beds}.
\newblock \bibinfo{journal}{Ind. Eng. Chem. Res.} \bibinfo{volume}{57},
  \bibinfo{pages}{3002--3013. \url{https://doi.org/10.1021/acs.iecr.7b04638}}.
\bibitem[{Di~Felice(1994)}]{difelice1994voidage}
\bibinfo{author}{Di~Felice, R.}, \bibinfo{year}{1994}.
\newblock \bibinfo{title}{The voidage function for fluid--particle interaction
  systems}.
\newblock \bibinfo{journal}{Int. J. Multiph. Flow} \bibinfo{volume}{20},
  \bibinfo{pages}{153--159.
  \url{https://doi.org/10.1016/0301--9322(94)90011--6}}.
\bibitem[{Ergun(1952)}]{ergun1952}
\bibinfo{author}{Ergun, S.}, \bibinfo{year}{1952}.
\newblock \bibinfo{title}{Fluid flow through packed columns}.
\newblock \bibinfo{journal}{Chem. Eng. Prog.} \bibinfo{volume}{48},
  \bibinfo{pages}{89--94}.
\bibitem[{Feng et~al.(2017)Feng, Li, Cheng, Yang and Fang}]{feng2017influence}
\bibinfo{author}{Feng, R.}, \bibinfo{author}{Li, J.}, \bibinfo{author}{Cheng,
  Z.}, \bibinfo{author}{Yang, X.}, \bibinfo{author}{Fang, Y.},
  \bibinfo{year}{2017}.
\newblock \bibinfo{title}{Influence of particle size distribution on minimum
  fluidization velocity and bed expansion at elevated pressure}.
\newblock \bibinfo{journal}{Powder Technol.} \bibinfo{volume}{320},
  \bibinfo{pages}{27--36. \url{https://doi.org/10.1016/j.powtec.2017.07.024}}.
\bibitem[{Gao et~al.(2024)Gao, Theuerkauf, Pakseresht, Kellogg and
  Fan}]{gao2024modified}
\bibinfo{author}{Gao, S.}, \bibinfo{author}{Theuerkauf, J.},
  \bibinfo{author}{Pakseresht, P.}, \bibinfo{author}{Kellogg, K.},
  \bibinfo{author}{Fan, Y.}, \bibinfo{year}{2024}.
\newblock \bibinfo{title}{A modified {Ergun} equation for application in packed
  beds with bidisperse and polydisperse spherical particles}.
\newblock \bibinfo{journal}{Powder Technol.} \bibinfo{volume}{445},
  \bibinfo{pages}{120035. \url{https://doi.org/10.1016/j.powtec.2024.120035}}.
\bibitem[{Glicksman et~al.(1993)Glicksman, Hyre and
  Woloshun}]{glicksman1993simplified}
\bibinfo{author}{Glicksman, L.R.}, \bibinfo{author}{Hyre, M.},
  \bibinfo{author}{Woloshun, K.}, \bibinfo{year}{1993}.
\newblock \bibinfo{title}{Simplified scaling relationships for fluidized beds}.
\newblock \bibinfo{journal}{Powder Technol.} \bibinfo{volume}{77},
  \bibinfo{pages}{177--199.
  \url{https://doi.org/10.1016/0032--5910(93)80055--F}}.
\bibitem[{Hoorijani et~al.(2024)Hoorijani, Esgandari, Zarghami,
  Sotudeh-Gharebagh and Mostoufi}]{hoorijani2024comparative}
\bibinfo{author}{Hoorijani, H.}, \bibinfo{author}{Esgandari, B.},
  \bibinfo{author}{Zarghami, R.}, \bibinfo{author}{Sotudeh-Gharebagh, R.},
  \bibinfo{author}{Mostoufi, N.}, \bibinfo{year}{2024}.
\newblock \bibinfo{title}{Comparative {CFD--DEM} study of flow regimes in
  spout-fluid beds}.
\newblock \bibinfo{journal}{Particuology} \bibinfo{volume}{85},
  \bibinfo{pages}{323--334.
  \url{https://doi.org/10.1016/j.partic.2023.07.011}}.
\bibitem[{Islam and Nguyen(2021a)}]{islam2021bed}
\bibinfo{author}{Islam, M.T.}, \bibinfo{author}{Nguyen, A.V.},
  \bibinfo{year}{2021}a.
\newblock \bibinfo{title}{Bed expansion and gas holdup characteristics of
  bubble--assisted fluidization of liquid--particle suspensions in a hydrofloat
  cell}.
\newblock \bibinfo{journal}{Miner. Eng.} \bibinfo{volume}{160},
  \bibinfo{pages}{106678. \url{https://doi.org/10.1016/j.mineng.2020.106678}}.
\bibitem[{Islam and Nguyen(2021b)}]{islam2021effect}
\bibinfo{author}{Islam, M.T.}, \bibinfo{author}{Nguyen, A.V.},
  \bibinfo{year}{2021}b.
\newblock \bibinfo{title}{Effect of particle size and shape on liquid--solid
  fluidization in a hydrofloat cell}.
\newblock \bibinfo{journal}{Powder Technol.} \bibinfo{volume}{379},
  \bibinfo{pages}{560--575.
  \url{https://doi.org/10.1016/j.powtec.2020.10.080}}.
\bibitem[{Islam and Nguyen(2023)}]{islam2023liquid}
\bibinfo{author}{Islam, M.T.}, \bibinfo{author}{Nguyen, A.V.},
  \bibinfo{year}{2023}.
\newblock \bibinfo{title}{Liquid-assisted irregular coarse particle
  fluidization in a fluidized bed flotation cell: Bed of low-density versus
  high-density particles}.
\newblock \bibinfo{journal}{Miner. Eng.} \bibinfo{volume}{201},
  \bibinfo{pages}{108153. \url{https://doi.org/10.1016/j.mineng.2023.108153}}.
\bibitem[{Khan et~al.(2016)Khan, Mitra, Ghatage, Peng, Doroodchi, Moghtaderi,
  Joshi and Evans}]{khan2016pressure}
\bibinfo{author}{Khan, M.S.}, \bibinfo{author}{Mitra, S.},
  \bibinfo{author}{Ghatage, S.}, \bibinfo{author}{Peng, Z.},
  \bibinfo{author}{Doroodchi, E.}, \bibinfo{author}{Moghtaderi, B.},
  \bibinfo{author}{Joshi, J.B.}, \bibinfo{author}{Evans, G.M.},
  \bibinfo{year}{2016}.
\newblock \bibinfo{title}{Pressure drop and voidage measurement in
  solid--liquid fluidized bed: Experimental, mathematical and computational
  study}.
\newblock \bibinfo{journal}{Chemeca} , \bibinfo{pages}{1019--1030.}
\bibitem[{Koekemoer and Luckos(2015)}]{koekemoer2015effect}
\bibinfo{author}{Koekemoer, A.}, \bibinfo{author}{Luckos, A.},
  \bibinfo{year}{2015}.
\newblock \bibinfo{title}{Effect of material type and particle size
  distribution on pressure drop in packed beds of large particles: Extending
  the {Ergun} equation}.
\newblock \bibinfo{journal}{Fuel} \bibinfo{volume}{158},
  \bibinfo{pages}{232--238. \url{https://doi.org/10.1016/j.fuel.2015.05.036}}.
\bibitem[{Kunii and Levenspiel(1991)}]{kunii1991fluidization}
\bibinfo{author}{Kunii, D.}, \bibinfo{author}{Levenspiel, O.},
  \bibinfo{year}{1991}.
\newblock \bibinfo{title}{Fluidization engineering}.
\newblock \bibinfo{edition}{2nd} ed.,
  \bibinfo{publisher}{Butterworth-Heinemann}.
\newblock \bibinfo{note}{\url{https://doi.org/10.1016/C2009-0-24190-0}}.
\bibitem[{Lacey(1954)}]{lacey1954developments}
\bibinfo{author}{Lacey, P.M.C.}, \bibinfo{year}{1954}.
\newblock \bibinfo{title}{Developments in the theory of particle mixing}.
\newblock \bibinfo{journal}{J. Appl. Chem.} \bibinfo{volume}{4},
  \bibinfo{pages}{257--268. \url{https://doi.org/10.1002/jctb.5010040504}}.
\bibitem[{Liu et~al.(2016)Liu, Yu, Lu, Wang, Liao and Hao}]{liu2016cfd}
\bibinfo{author}{Liu, G.}, \bibinfo{author}{Yu, F.}, \bibinfo{author}{Lu, H.},
  \bibinfo{author}{Wang, S.}, \bibinfo{author}{Liao, P.}, \bibinfo{author}{Hao,
  Z.}, \bibinfo{year}{2016}.
\newblock \bibinfo{title}{{CFD--DEM} simulation of liquid--solid fluidized bed
  with dynamic restitution coefficient}.
\newblock \bibinfo{journal}{Powder Technol.} \bibinfo{volume}{304},
  \bibinfo{pages}{186--197.
  \url{https://doi.org/10.1016/j.powtec.2016.08.058}}.
\bibitem[{Lu et~al.(2015)Lu, Huang, Zheng and Jing}]{lu2015flow}
\bibinfo{author}{Lu, Y.}, \bibinfo{author}{Huang, J.}, \bibinfo{author}{Zheng,
  P.}, \bibinfo{author}{Jing, D.}, \bibinfo{year}{2015}.
\newblock \bibinfo{title}{Flow structure and bubble dynamics in supercritical
  water fluidized bed and gas fluidized bed: A comparative study}.
\newblock \bibinfo{journal}{Int. J. Multiph. Flow} \bibinfo{volume}{73},
  \bibinfo{pages}{130--141.
  \url{https://doi.org/10.1016/j.ijmultiphaseflow.2015.03.011}}.
\bibitem[{Luo et~al.(2015)Luo, Wu, Yang and Fan}]{luo2015cfd}
\bibinfo{author}{Luo, K.}, \bibinfo{author}{Wu, F.}, \bibinfo{author}{Yang,
  S.}, \bibinfo{author}{Fan, J.}, \bibinfo{year}{2015}.
\newblock \bibinfo{title}{{CFD--DEM} study of mixing and dispersion behaviors
  of solid phase in a bubbling fluidized bed}.
\newblock \bibinfo{journal}{Powder Technol.} \bibinfo{volume}{274},
  \bibinfo{pages}{482--493.
  \url{https://doi.org/10.1016/j.powtec.2015.01.046}}.
\bibitem[{Ma et~al.(2017)Ma, Xu and Zhao}]{ma2017cfd}
\bibinfo{author}{Ma, H.}, \bibinfo{author}{Xu, L.}, \bibinfo{author}{Zhao, Y.},
  \bibinfo{year}{2017}.
\newblock \bibinfo{title}{{CFD--DEM} simulation of fluidization of rod-like
  particles in a fluidized bed}.
\newblock \bibinfo{journal}{Powder Technol.} \bibinfo{volume}{314},
  \bibinfo{pages}{355--366.
  \url{https://doi.org/10.1016/j.powtec.2016.12.008}}.
\bibitem[{Marchelli and Di~Felice(2023)}]{marchelli2023drag}
\bibinfo{author}{Marchelli, F.}, \bibinfo{author}{Di~Felice, R.},
  \bibinfo{year}{2023}.
\newblock \bibinfo{title}{An experimental assessment of fluid--solid drag
  models based on the pressure drop in bidisperse fixed beds}.
\newblock \bibinfo{journal}{Int. J. Multiph. Flow} \bibinfo{volume}{166},
  \bibinfo{pages}{104513.
  \url{https://doi.org/10.1016/j.ijmultiphaseflow.2023.104513}}.
\bibitem[{Moliner et~al.(2019)Moliner, Marchelli, Spanachi, Martinez-Felipe,
  Bosio and Arato}]{moliner2019cfd}
\bibinfo{author}{Moliner, C.}, \bibinfo{author}{Marchelli, F.},
  \bibinfo{author}{Spanachi, N.}, \bibinfo{author}{Martinez-Felipe, A.},
  \bibinfo{author}{Bosio, B.}, \bibinfo{author}{Arato, E.},
  \bibinfo{year}{2019}.
\newblock \bibinfo{title}{{CFD} simulation of a spouted bed: Comparison between
  the discrete element method {(DEM)} and the two fluid model {(TFM)}}.
\newblock \bibinfo{journal}{Chem. Eng. J.} \bibinfo{volume}{377},
  \bibinfo{pages}{120466. \url{https://doi.org/10.1016/j.cej.2018.11.164}}.
\bibitem[{Peng et~al.(2022)Peng, Sun, Han, Xie and Xiao}]{peng2022experimental}
\bibinfo{author}{Peng, J.}, \bibinfo{author}{Sun, W.}, \bibinfo{author}{Han,
  H.}, \bibinfo{author}{Xie, L.}, \bibinfo{author}{Xiao, Y.},
  \bibinfo{year}{2022}.
\newblock \bibinfo{title}{Experimental and numerical simulation study on the
  hydrodynamic characteristics of spherical and irregular-shaped particles in a
  {3D} liquid-fluidized bed}.
\newblock \bibinfo{journal}{Korean J. Chem. Eng.} \bibinfo{volume}{39},
  \bibinfo{pages}{3165--3176.
  \url{https://doi.org/10.1007/s11814--022--1234--9}}.
\bibitem[{Peng et~al.(2016)Peng, Alghamdi, Moghtaderi and
  Doroodchi}]{peng2016cfd}
\bibinfo{author}{Peng, Z.}, \bibinfo{author}{Alghamdi, Y.A.},
  \bibinfo{author}{Moghtaderi, B.}, \bibinfo{author}{Doroodchi, E.},
  \bibinfo{year}{2016}.
\newblock \bibinfo{title}{{CFD--DEM} investigation of transition from
  segregation to mixing of binary solids in gas fluidised beds}.
\newblock \bibinfo{journal}{Adv. Powder Technol.} \bibinfo{volume}{27},
  \bibinfo{pages}{2342--2353. \url{https://doi.org/10.1016/j.apt.2016.08.024}}.
\bibitem[{Peng et~al.(2014a)Peng, Doroodchi, Luo and
  Moghtaderi}]{peng2014influence}
\bibinfo{author}{Peng, Z.}, \bibinfo{author}{Doroodchi, E.},
  \bibinfo{author}{Luo, C.}, \bibinfo{author}{Moghtaderi, B.},
  \bibinfo{year}{2014}a.
\newblock \bibinfo{title}{Influence of void fraction calculation on fidelity of
  {CFD--DEM} simulation of gas--solid bubbling fluidized beds}.
\newblock \bibinfo{journal}{AIChE J.} \bibinfo{volume}{60},
  \bibinfo{pages}{2000--2018. \url{https://doi.org/10.1002/aic.14421}}.
\bibitem[{Peng et~al.(2014b)Peng, Ghatage, Doroodchi, Joshi, Evans and
  Moghtaderi}]{peng2014forces}
\bibinfo{author}{Peng, Z.}, \bibinfo{author}{Ghatage, S.V.},
  \bibinfo{author}{Doroodchi, E.}, \bibinfo{author}{Joshi, J.B.},
  \bibinfo{author}{Evans, G.M.}, \bibinfo{author}{Moghtaderi, B.},
  \bibinfo{year}{2014}b.
\newblock \bibinfo{title}{Forces acting on a single introduced particle in a
  solid--liquid fluidised bed}.
\newblock \bibinfo{journal}{Chem. Eng. Sci.} \bibinfo{volume}{116},
  \bibinfo{pages}{49--70. \url{https://doi.org/10.1016/j.ces.2014.04.040}}.
\bibitem[{Puhan et~al.(2021)Puhan, Awasthi, Mukherjee and Atta}]{puhan2021cfd}
\bibinfo{author}{Puhan, P.}, \bibinfo{author}{Awasthi, A.},
  \bibinfo{author}{Mukherjee, A.K.}, \bibinfo{author}{Atta, A.},
  \bibinfo{year}{2021}.
\newblock \bibinfo{title}{{CFD} modeling of segregation in binary solid--liquid
  fluidized beds: Influence of liquid viscosity and density}.
\newblock \bibinfo{journal}{Chem. Eng. Sci.} \bibinfo{volume}{246},
  \bibinfo{pages}{116965. \url{https://doi.org/10.1016/j.ces.2021.116965}}.
\bibitem[{Sakai et~al.(2014)Sakai, Abe, Shigeto, Mizutani, Takahashi, Vir{\'e},
  Percival, Xiang and Pain}]{sakai2014verification}
\bibinfo{author}{Sakai, M.}, \bibinfo{author}{Abe, M.},
  \bibinfo{author}{Shigeto, Y.}, \bibinfo{author}{Mizutani, S.},
  \bibinfo{author}{Takahashi, H.}, \bibinfo{author}{Vir{\'e}, A.},
  \bibinfo{author}{Percival, J.R.}, \bibinfo{author}{Xiang, J.},
  \bibinfo{author}{Pain, C.C.}, \bibinfo{year}{2014}.
\newblock \bibinfo{title}{Verification and validation of a coarse grain model
  of the {DEM} in a bubbling fluidized bed}.
\newblock \bibinfo{journal}{Chem. Eng. J.} \bibinfo{volume}{244},
  \bibinfo{pages}{33--43. \url{https://doi.org/10.1016/j.cej.2014.01.029}}.
\bibitem[{Thornton(2015)}]{thornton2015granular}
\bibinfo{author}{Thornton, C.}, \bibinfo{year}{2015}.
\newblock \bibinfo{title}{Granular dynamics, contact mechanics and particle
  system simulations: A {DEM} study}.
\newblock Particle Technology Series, \bibinfo{publisher}{Springer Cham}.
\newblock \bibinfo{note}{\url{https://doi.org/10.1007/978-3-319-18711-2}}.
\bibitem[{Wang and Shen(2024)}]{wang2024cfd}
\bibinfo{author}{Wang, S.}, \bibinfo{author}{Shen, Y.}, \bibinfo{year}{2024}.
\newblock \bibinfo{title}{{CFD--DEM} study of thermal behaviours of chip-like
  particles flow in a fluidized bed}.
\newblock \bibinfo{journal}{Powder Technol.} \bibinfo{volume}{445},
  \bibinfo{pages}{120066. \url{https://doi.org/10.1016/j.powtec.2024.120066}}.
\bibitem[{Wen and Yu(1966)}]{wen1966generalized}
\bibinfo{author}{Wen, C.Y.}, \bibinfo{author}{Yu, Y.H.}, \bibinfo{year}{1966}.
\newblock \bibinfo{title}{A generalized method for predicting the minimum
  fluidization velocity}.
\newblock \bibinfo{journal}{AIChE J.} \bibinfo{volume}{12},
  \bibinfo{pages}{610--612. \url{https://doi.org/10.1002/aic.690120343}}.
\bibitem[{Xie et~al.(2025)Xie, Wang, Shao, Chen, Ding and Ma}]{xie2025cfd}
\bibinfo{author}{Xie, L.}, \bibinfo{author}{Wang, S.}, \bibinfo{author}{Shao,
  B.}, \bibinfo{author}{Chen, X.}, \bibinfo{author}{Ding, N.},
  \bibinfo{author}{Ma, Y.}, \bibinfo{year}{2025}.
\newblock \bibinfo{title}{{CFD--DEM} study on mixing and segregation
  characteristics for binary column-shape particles in a liquid--solid
  fluidized bed}.
\newblock \bibinfo{journal}{Adv. Powder Technol.} \bibinfo{volume}{36},
  \bibinfo{pages}{104788. \url{https://doi.org/10.1016/j.apt.2025.104788}}.
\bibitem[{Xie et~al.(2021)Xie, Wang and Shen}]{xie2021cfd}
\bibinfo{author}{Xie, Z.}, \bibinfo{author}{Wang, S.}, \bibinfo{author}{Shen,
  Y.}, \bibinfo{year}{2021}.
\newblock \bibinfo{title}{{CFD--DEM} study of segregation and mixing
  characteristics under a bi-disperse solid--liquid fluidised bed}.
\newblock \bibinfo{journal}{Adv. Powder Technol.} \bibinfo{volume}{32},
  \bibinfo{pages}{4078--4095. \url{https://doi.org/10.1016/j.apt.2021.09.012}}.
\bibitem[{Xin et~al.(2025)Xin, Hao, Pan, Zhang and Kong}]{xin2025fluid}
\bibinfo{author}{Xin, H.}, \bibinfo{author}{Hao, H.}, \bibinfo{author}{Pan,
  Z.}, \bibinfo{author}{Zhang, D.}, \bibinfo{author}{Kong, B.},
  \bibinfo{year}{2025}.
\newblock \bibinfo{title}{Fluid and particle velocity fluctuation measurements
  in a mono- and bidisperse liquid--solid fluidized bed}.
\newblock \bibinfo{journal}{Chem. Eng. Sci.} , \bibinfo{pages}{122753.
  \url{https://doi.org/10.1016/j.ces.2025.122753}}.
\bibitem[{Zhang et~al.(2021)Zhang, Huang, An, Yu and Xie}]{zhang2021numerical}
\bibinfo{author}{Zhang, H.}, \bibinfo{author}{Huang, Y.}, \bibinfo{author}{An,
  X.}, \bibinfo{author}{Yu, A.}, \bibinfo{author}{Xie, J.},
  \bibinfo{year}{2021}.
\newblock \bibinfo{title}{Numerical prediction on the minimum fluidization
  velocity of a supercritical water fluidized bed reactor: Effect of particle
  size distributions}.
\newblock \bibinfo{journal}{Powder Technol.} \bibinfo{volume}{389},
  \bibinfo{pages}{119--130.
  \url{https://doi.org/10.1016/j.powtec.2021.05.015}}.
\bibitem[{Zhou et~al.(2022)Zhou, Lv, Bai, Han, Wang, Shi and
  Huang}]{zhou2022cfd}
\bibinfo{author}{Zhou, L.}, \bibinfo{author}{Lv, W.}, \bibinfo{author}{Bai,
  L.}, \bibinfo{author}{Han, Y.}, \bibinfo{author}{Wang, J.},
  \bibinfo{author}{Shi, W.}, \bibinfo{author}{Huang, G.}, \bibinfo{year}{2022}.
\newblock \bibinfo{title}{{CFD--DEM} study of gas--solid flow characteristics
  in a fluidized bed with different diameter of coarse particles}.
\newblock \bibinfo{journal}{Energy Rep.} \bibinfo{volume}{8},
  \bibinfo{pages}{2376--2388.
  \url{https://doi.org/10.1016/j.egyr.2022.01.174}}.

\end{thebibliography}

\end{document}